%% file: main.tex
\documentclass[10pt]{article} 
\usepackage[accepted]{tmlr}

\usepackage[T1]{fontenc}

\usepackage{hyperref}
\usepackage{url}

\usepackage{graphicx}

\usepackage{booktabs}
\usepackage{multirow}
\usepackage{float}
\usepackage[normalem]{ulem}

\usepackage{xcolor}
\usepackage{colortbl}
\usepackage{array}
\usepackage{ragged2e}

\usepackage[all]{nowidow}

\usepackage{enumitem}

\usepackage{hyphenat}

\input{sections/_authors}

\def\month{08}  
\def\year{2026} 
\def\openreview{\url{https://openreview.net/forum?id=ypRg1TvQaM}} 

\begin{document}

\maketitle

\input{sections/abstract}

\input{sections/intro}

\input{sections/difficulty}
\input{sections/preference}
\input{sections/clustering}

\input{sections/rl}
\input{sections/uncertainty}
\input{sections/cascades}

\input{sections/multimodal}

\input{sections/evaluation}

\input{sections/framework}
\input{sections/future}

\input{sections/ethics}
\input{sections/acks}

\input{tables/comparison}

\newpage
\bibliography{paperpile}
\bibliographystyle{tmlr}

\end{document}

%% file: sections/_authors.tex
\title{Dynamic Model Routing and Cascading for \\ Efficient LLM Inference: A Survey}

\author{\name Yasmin Moslem
      \email yasmin.moslem@adaptcentre.ie \\
      \addr ADAPT Centre \\
      School of Computer Science \& Statistics \\
      Trinity College Dublin \\
      Dublin, Ireland
      \AND
      \name John D. Kelleher 
      \email john.kelleher@adaptcentre.ie \\
      \addr ADAPT Centre \\
      School of Computer Science \& Statistics \\
      Trinity College Dublin \\
      Dublin, Ireland
      }

%% file: sections/abstract.tex
\begin{abstract}
\nohyphens{
The rapid growth of large language models (LLMs) with diverse capabilities, costs, and domains has created a critical need for intelligent model selection at inference time.
While smaller models suffice for routine queries, complex tasks demand more capable models. However, static model deployment does not account for the complexity and domain of incoming queries, leading to suboptimal performance and increased costs.
Dynamic routing systems that adaptively select models based on query characteristics have emerged as a solution to this challenge.

This survey provides a systematic analysis of multi-LLM routing and cascading approaches, focusing on systems that route queries across a pool of independently trained LLMs at inference time.
We cover diverse routing paradigms, including query difficulty, human preferences, clustering, uncertainty quantification, reinforcement learning, multimodality, and cascading.
For each paradigm, we analyze representative methods and examine key trade-offs. Beyond taxonomy, we introduce a conceptual framework that characterizes routing systems along three dimensions: when decisions are made, what information is used, and how they are computed.
This perspective highlights that practical systems are often compositional, integrating multiple paradigms under operational constraints.

Our analysis demonstrates that effective multi-LLM routing requires balancing competing objectives.~%
Choosing the optimal routing strategy depends on deployment and computational constraints.~%
Well-designed routing systems can outperform even the most powerful individual models by strategically leveraging specialized capabilities across models while maximizing efficiency gains.
Meanwhile, open challenges remain in developing and evaluating routing mechanisms that generalize across diverse architectures, modalities, and applications.
}

\end{abstract}

%% file: sections/intro.tex
\newpage
\section{Introduction}
\label{sec:introduction}

\subsection{Problem and Motivation}

Production deployment of large language models faces a fundamental cost-performance dilemma. Queries vary substantially in complexity, from simple factual questions to complex multi-step reasoning problems. When a single model handles all requests, simple queries consume unnecessary resources when routed to powerful models, while complex queries may exceed the capabilities of smaller models.

Recent advances in multi-LLM deployment have introduced dynamic routing systems that address this challenge. These systems analyze each query and select from a pool of models with different capabilities, costs, and specializations. By matching computational resources to query requirements, adaptive routing can reduce costs while maintaining or improving output quality.

\subsection{Routing and Cascading}

This survey covers two complementary approaches to adaptive model selection at inference time:

\textbf{Model routing} analyzes each input and selects the most appropriate model based on query characteristics. The router makes a single decision, mapping the query to one model from the available pool.

\textbf{Model cascading} operates sequentially, first attempting inference with smaller, faster models and escalating to larger, more capable models only when the initial response is deemed insufficient based on quality estimation.

Both approaches aim to optimize the performance-cost trade-off by matching queries to appropriate models. Production systems often combine routing and cascading strategies to maximize efficiency.

\subsection{Scope and Organization}
\label{sec:scope}

This survey focuses on routing between independently trained LLMs at inference time, where the goal is to select the most suitable model from a discrete pool given a query.
In all cases, the routing mechanism is treated as a separate orchestration layer, distinct from the candidate models it selects among.
We organize methods into six main paradigms based on their routing strategy, summarized in Table~\ref{tab:methods}, while acknowledging that methods may span multiple categories:

\begin{itemize}
\item \textbf{Difficulty-aware routing} (Section \ref{sec:difficulty}): Routes based on estimated query complexity 
\item \textbf{Human preference-aligned routing} (Section \ref{sec:preference}): Leverages preference data from human feedback
\item \textbf{Clustering-based routing} (Section \ref{sec:clustering}): Groups similar queries using unsupervised learning
\item \textbf{Reinforcement learning routing} (Section \ref{sec:rl}): Learns routing decisions through reinforcement learning, including policy optimization (e.g.\ PPO, GRPO) and bandit methods
\item \textbf{Uncertainty-based routing} (Section \ref{sec:uncertainty}): Routes based on model confidence estimates
\item \textbf{Cascading} (Section \ref{sec:cascades}): Sequential multi-model approaches
\end{itemize}

Additionally, we briefly explore research on \textbf{routing of multimodal models} (Section \ref{sec:multimodal}), and discuss \textbf{evaluation approaches}, including benchmarks and metrics (Section \ref{sec:evaluation}). We then synthesize the surveyed methods through a \textbf{multidimensional framework} in Section \ref{sec:framework}. Finally, we discuss open challenges and \textbf{future directions} (Section \ref{sec:future}), providing a foundation for advancing research in efficient multi-LLM deployment and evaluation in production.

\newpage
\subsection{Conceptual Design Space for LLM Routing}
\label{sec:intro-framework}

The paradigms we cover in this survey (cf.~Section \ref{sec:scope}) provide a useful foundation for organizing and understanding the literature. In practice, real-world systems draw on more than one of these paradigms simultaneously. To complement the paradigm-based organization, routing approaches can be categorized through a multidimensional conceptual framework.

\paragraph{When the routing decision is made.} Routing systems can rely on either pre-generation or post-generation decisions, or they can adopt a multi-stage process. Pre-generation routing selects a model before generating any output, relying entirely on properties of the incoming query, while post-generation routing systems make their decision after an initial response has been produced, using output quality or confidence as the primary signal. Instead of making a one-time decision, some approaches embed routing within a sequential multi-stage process, escalating across models based on response quality. Multi-stage systems are not mutually exclusive with pre- or post-generation routing; they typically combine both within a sequential escalation structure.

\paragraph{What information the routing mechanism uses.} 
Routing systems differ substantially in the richness of their input signals. The simplest approaches operate on the query alone, using lexical or semantic features to characterize the request. More informed systems additionally 
incorporate metadata about available models to guide selection, such as cost, latency, or domain specialization. Post-generation approaches further extend this by incorporating response-level signals, such as confidence scores, token probabilities, or verifier outputs. Some systems also accumulate external feedback over time, adapting their routing behaviour based on user interactions or downstream task performance.

\paragraph{How the decision is computed.} 
Routing decisions vary considerably in their computational complexity. At one end, simple threshold rules or cost-based heuristics require no training and can be applied directly at inference time. At the other end, supervised classifiers trained on historical performance data learn to predict which model is likely to handle a given query best. More sophisticated approaches either employ bandit algorithms that update routing behaviour through ongoing interaction at deployment time, or reinforcement learning methods such as PPO and GRPO that optimize a routing policy during training. In practice, many systems combine these mechanisms, for example using a classifier to make an initial routing decision and a threshold rule to trigger escalation within a cascade.

These dimensions are not independent of the paradigms. Table~\ref{tab:framework} maps representative methods from each of the six paradigms to these dimensions.
In this sense, clustering and difficulty-aware methods are generally pre-generation approaches that operate on query-level signals, while uncertainty-based methods and cascades typically involve post-generation decisions on response-level signals. Interestingly, cascading approaches tend to employ layers of these individual techniques for diverse objectives, ranging from balancing performance-cost trade-offs to implementing safety measures on input queries and output generations. This multidimensional view helps reveal where systems genuinely overlap, as real-world deployments rarely conform to a single paradigm. Production systems typically combine multiple routing strategies, adapting their behaviour to the demands of diverse queries, operational constraints, and evolving user needs. This framework therefore offers a unifying lens that complements the paradigm-based organization of the sections that follow.
Section~\ref{sec:framework} revisits this framework after the paradigm-based survey, examining how production systems compose across these dimensions and identifying structural gaps in the literature.

\input{tables/framework}

%% file: tables/framework.tex
%

\definecolor{diffcol}{HTML}{DBEAFE}   
\definecolor{prefcol}{HTML}{DCFCE7}   
\definecolor{clustcol}{HTML}{FEF9C3}  
\definecolor{rlcol}{HTML}{FFEDD5}     
\definecolor{unccol}{HTML}{F3E8FF}    
\definecolor{casccol}{HTML}{FFE4E6}   

\definecolor{diffcell}{HTML}{B3D8FD}  
\definecolor{prefcell}{HTML}{6EE7A0}  
\definecolor{clustcell}{HTML}{FDDD76} 
\definecolor{rlcell}{HTML}{FBAA64}    
\definecolor{unccell}{HTML}{D3A4FD}   
\definecolor{casccell}{HTML}{FC8FA0}  

\newcommand{\Yd}{\cellcolor{diffcell}}   
\newcommand{\Yp}{\cellcolor{prefcell}}   
\newcommand{\Yc}{\cellcolor{clustcell}}  
\newcommand{\Yr}{\cellcolor{rlcell}}     
\newcommand{\Yu}{\cellcolor{unccell}}    
\newcommand{\Yk}{\cellcolor{casccell}}   

\begin{table*}[p]
\centering
\scriptsize
\setlength{\tabcolsep}{3pt}
\renewcommand{\arraystretch}{1.25}
%
%
\begin{tabular}{>{\raggedright\arraybackslash}p{2.6cm} cccc cccc cccc}
\toprule
\multirow{2}{*}{\textbf{Method}}
  & \multicolumn{4}{c}{\textbf{When}}
  & \multicolumn{4}{c}{\textbf{What signals}}
  & \multicolumn{4}{c}{\textbf{How computed}} \\
\cmidrule(lr){2-5}\cmidrule(lr){6-9}\cmidrule(lr){10-13}
& \rotatebox{65}{\small Pre-gen.}
& \rotatebox{65}{\small Post-gen.}
& \rotatebox{65}{\small Multi-stage}
& \rotatebox{65}{\small Online}
& \rotatebox{65}{\small Query}
& \rotatebox{65}{\small Model meta}
& \rotatebox{65}{\small Response}
& \rotatebox{65}{\small Feedback}
& \rotatebox{65}{\small Heuristic}
& \rotatebox{65}{\small Supervised}
& \rotatebox{65}{\small Bandit}
& \rotatebox{65}{\small RL policy} \\
\midrule
\multicolumn{13}{l}{\cellcolor{diffcol}\textit{Difficulty-aware}} \\
\rowcolor{diffcol}
BEST-Route            & \Yd &      &      &      & \Yd & \Yd &      &      & \Yd & \Yd &      &      \\
\rowcolor{diffcol}
vLLM Router  & \Yd &      &      &      & \Yd &      &      &      &      & \Yd &      &      \\
\rowcolor{diffcol}
RouteLMT & \Yd &      &      &      & \Yd &      &      &      &      & \Yd &      &      \\
\rowcolor{diffcol}
EmbedLLM              & \Yd &      &      &      & \Yd & \Yd &      &      &      & \Yd &      &      \\
\rowcolor{diffcol}
ICL-Router & \Yd &      &      &      & \Yd & \Yd &      &      &      & \Yd &      &      \\
\rowcolor{diffcol}
GraphRouter           & \Yd &      &      &      & \Yd & \Yd &      &      &      & \Yd &      &      \\
\rowcolor{diffcol}
IRT-Router & \Yd &      &      &      & \Yd & \Yd &      &      &      & \Yd &      &      \\
\midrule
\multicolumn{13}{l}{\cellcolor{prefcol}\textit{Preference-aligned}} \\
\rowcolor{prefcol}
RouteLLM                 & \Yp &      &      &      & \Yp &      &      & \Yp &      & \Yp &      &      \\
\rowcolor{prefcol}
Hybrid LLM             & \Yp &      &      &      & \Yp &      &      & \Yp &      & \Yp &      &      \\
\rowcolor{prefcol}
Arch-Router          & \Yp &      &      &      & \Yp &      &      &      &      & \Yp &      &      \\
\rowcolor{prefcol}
P2L       & \Yp &      &      &      & \Yp &      &      & \Yp &      & \Yp &      &      \\
\rowcolor{prefcol}
Eagle                      & \Yp &      &      &      &      &      &      & \Yp & \Yp &      &      &      \\
\rowcolor{prefcol}
Zooter                      & \Yp &      &      &      & \Yp &      &      &      &      & \Yp &      &      \\
\midrule
\multicolumn{13}{l}{\cellcolor{clustcol}\textit{Clustering-based}} \\
\rowcolor{clustcol}
UniRoute          & \Yc &      &      &      & \Yc & \Yc &      &      & \Yc & \Yc &      &      \\
\rowcolor{clustcol}
Avengers-Pro       & \Yc &      &      &      & \Yc & \Yc &      &      & \Yc & \Yc &      &      \\
\midrule
\multicolumn{13}{l}{\cellcolor{rlcol}\textit{Reinforcement Learning}} \\
\rowcolor{rlcol}
Router-R1             & \Yr & \Yr & \Yr &      & \Yr & \Yr &      &      &      & \Yr &      & \Yr \\
\rowcolor{rlcol}
R2-Reasoner          & \Yr & \Yr & \Yr &      & \Yr & \Yr &      &      &      & \Yr &      & \Yr \\
\rowcolor{rlcol}
SCOPE                 & \Yr &      &      &      & \Yr & \Yr &      &      & \Yr & \Yr &      & \Yr \\
\rowcolor{rlcol}
MetaLLM                & \Yr &      &      & \Yr & \Yr &      &      & \Yr &      &      & \Yr &      \\
\rowcolor{rlcol}
MixLLM                    & \Yr &      &      & \Yr & \Yr & \Yr &      & \Yr &      &      & \Yr &      \\
\rowcolor{rlcol}
PILOT                     & \Yr &      &      & \Yr & \Yr & \Yr &      & \Yr &      &      & \Yr &      \\
\rowcolor{rlcol}
GreenServ  & \Yr &      &      & \Yr & \Yr &      &      & \Yr & \Yr & \Yr & \Yr &      \\\rowcolor{rlcol}
Dueling Feedback  & \Yr &      &      & \Yr & \Yr & \Yr &      & \Yr &      & \Yr & \Yr &      \\
\rowcolor{rlcol}
TI-UCB                    & \Yr &      &      & \Yr &      &      &      & \Yr &      &      & \Yr &      \\
\midrule
\multicolumn{13}{l}{\cellcolor{unccol}\textit{Uncertainty-based}} \\
\rowcolor{unccol}
CP-Router                 &      & \Yu &      &      &      &      & \Yu &      & \Yu &      &      &      \\
\rowcolor{unccol}
Probe-based UQ                 &      & \Yu &      &      &      &      & \Yu &      &      & \Yu &      &      \\
\midrule
\multicolumn{13}{l}{\cellcolor{casccol}\textit{Cascading}} \\
\rowcolor{casccol}
FrugalGPT              & \Yk & \Yk & \Yk &      & \Yk & \Yk & \Yk &      & \Yk & \Yk &      &      \\
\rowcolor{casccol}
Cascade Routing  & \Yk & \Yk & \Yk &      & \Yk & \Yk & \Yk &      & \Yk & \Yk &      &      \\
\rowcolor{casccol}
CRE-Router       & \Yk & \Yk & \Yk &      & \Yk & \Yk & \Yk &      & \Yk & \Yk &      &      \\
\rowcolor{casccol}
AutoMix               &      & \Yk & \Yk &      &      &      & \Yk &      & \Yk &      &      &      \\
\rowcolor{casccol}
Self-REF               &      & \Yk & \Yk &      &      &      & \Yk &      & \Yk & \Yk &      &      \\
\rowcolor{casccol}
LLM-Blender          &      & \Yk & \Yk &      &      &      & \Yk &      &      & \Yk &      &      \\
\bottomrule
\end{tabular}
\caption{Design-space matrix mapping representative routing and cascading methods to the survey's conceptual framework along three axes: \emph{when} routing decisions are made (Pre-gen., Post-gen., Multi-stage, Online/adaptive), \emph{what} signals are used (Query, Model metadata, Response, Feedback), and \emph{how} decisions are computed (Heuristic, Supervised, Bandit, RL policy). Shaded cells indicate the property applies; cell and row colour identify the paradigm group. Multi-stage denotes a sequential escalation structure and is not mutually exclusive with Pre-gen.\ and Post-gen., which describe the types of decisions made within that structure. Multiple filled cells in the \emph{how} group indicate that a method combines mechanisms, e.g.\ a learned classifier paired with a threshold rule. This framework is introduced in Section~\ref{sec:intro-framework} and applied in Section~\ref{sec:framework} to synthesize compositional patterns across surveyed methods and surface structural gaps in the literature.}
\label{tab:framework}
\end{table*}

%% file: sections/difficulty.tex
\section{Difficulty-aware Routing}
\label{sec:difficulty}

When an LLM is able to answer a query correctly, it is often because the query is relatively simple or closely aligned with the model's training data. On the contrary, more complex queries such as maths or sophisticated coding tasks may require advanced reasoning capabilities. Driven by this observation that some queries are inherently more ``difficult'' than others, researchers have explored the idea of routing queries to different LLMs based on their difficulty level.

Query difficulty can be assessed using various methods, including heuristic-based approaches (e.g. text length, word rarity, idiomatic language, syntactic complexity), learned classifiers, or even leveraging the LLMs themselves to estimate the complexity of a query, usually referred to as ``LLM as a judge'' \citep{Proietti2025-EstimatingMTDifficulty}. Once the difficulty of a query is estimated, it can be routed to an appropriate LLM. For instance, simpler queries can be directed to smaller, more cost-effective models, while complex queries can be sent to larger, more capable models. This optimizes resource allocation while maintaining performance. In this sense, several routing methods can fall under this dynamic model selection paradigm, tailored to query difficulty.
Implementations can prioritize different objectives, such as cost, latency, accuracy, or human preference.
Within the design-space framework (cf.~Table~\ref{tab:framework}), difficulty-aware methods are characteristically pre-generation approaches that operate on query-level signals, typically implemented as supervised classifiers or heuristics.

Among the work that can be classified under the difficulty-aware routing paradigm is BEST-Route \citep{Ding2025-BEST-Route} which dynamically allocates queries to different LLMs depending on their assessed difficulty. BEST-Route uses a multi-head router based on DeBERTa-v3-small \citep{He2020-DeBERTa,He2021-DeBERTaV3} to estimate query difficulty and select the optimal model and sampling strategy. At inference time, it employs best-of-$n$ sampling \citep{Stiennon2020-SummarizeHumanFeedback,Nakano2021-WebGPT-HumanFeedback} to enhance the performance of small models, selecting the best response out of $n$ generated responses, using a proxy reward model fine-tuned from OpenAssistant RM, which is based on a DeBERTa-v3-large.
If necessary, sampling is increased, or harder queries are routed to a larger model.
At inference time, users can set thresholds to balance cost and accuracy, where higher thresholds favour the reference model, improving quality at increased costs. Since multiple small-model and best-of-$n$ combinations can meet a given threshold, BEST-Route effectively selects among them using cost estimation to ensure high-quality responses at minimal cost.
HAPS \citep{Tian2026-HAPS} pursues a related idea of routing beyond model choice alone, but targets the model's parameters rather than a sampling strategy: a high-level router selects the LLM, and a low-level router, implemented as a parameter-generation network sharing parameters with the high-level router, generates input-conditioned LoRA adapter weights for the selected model. The two routers are trained jointly, so that model selection and parameter generation reinforce each other.

In the area of machine translation, \citet{Wu2025-CombiningBestTranslation} apply a similar intuition, proposing source-side routing policies that decide whether to use an NMT system or an LLM based on the complexity of the source sentence. RouteLMT \citep{Luo2026-RouteLMT} instead proposes an in-model router that probes the small translator's prompt-token representations via lightweight LoRA adaptation, without requiring external models or hypothesis decoding, identifying the marginal gain, i.e., the large model's improvement over the small model, as the optimal signal for routing decisions.

Similarly, vLLM Semantic Router \citep{Wang2025-vLLM-SemanticRouter} addresses the challenge of adaptive reasoning, where costly step-by-step inference is applied only when beneficial, while maintaining low latency and efficiency for straightforward queries. The system uses a ModernBERT-based classifier \citep{Warner2025-ModernBERT} to analyze query intent and complexity, routing queries that require reasoning to models with chain-of-thought capabilities while directing simpler queries to standard inference without reasoning. This reasoning-aware approach demonstrates that selective application of computational resources based on query complexity can simultaneously improve both accuracy and efficiency.

An early formulation of this idea recast router training as a set of binary correctness-classification problems, estimated in practice with a k-nearest-neighbour classifier \citep{Shnitzer2024-LLMRoutingBenchmarkDatasets}, a framing that later predictive-scoring routers such as EmbedLLM and ICL-Router build on. PORT \citep{Wu2025-PORT} scales this k-nearest-neighbour estimation to high-volume online serving, using approximate nearest-neighbour search over a historical query set to estimate each candidate model's performance and cost for an incoming query, without training a parametric router. It calibrates a set of routing weights once, from an initial slice of the query stream, and applies those weights to route the remaining queries under a per-model token budget.

In a relevant line of work, EmbedLLM \citep{Zhuang2024-EmbedLLM} employs matrix factorization to learn compact vector embeddings of LLMs, capturing model characteristics such as domain specialization. At routing time, these embeddings are used to predict which model will answer a given query correctly, effectively estimating query-model compatibility without requiring explicit difficulty labels. The approach enables efficient routing and benchmark accuracy prediction, though it requires retraining when incorporating new models.

RouterDC \citep{Chen2024-RouterDC} takes a contrastive-learning approach to the same correctness-prediction problem, training a shared encoder alongside per-LLM embedding vectors. A sample-LLM contrastive loss pulls each query toward the embeddings of LLMs that answer it correctly while pushing away those that do not; a complementary sample-sample contrastive loss groups semantically similar queries together, with groups defined by K-means clustering on the training set. At inference, routing reduces to selecting the LLM whose embedding is most similar to the query representation, a single forward pass with no LLM calls until the final selected model.

ICL-Router \citep{Wang2026-ICL-Router} represents model capabilities as compact in-context vectors, derived from a small set of query-performance pairs that reflect each model's capability profile. An LLM-based router is trained in two stages, first to align query embeddings with its semantic space, then to predict whether each model can correctly answer a given query using these vectors as in-context input. A key advantage is that new models can be integrated without retraining the router, by simply evaluating them on a representative query set to generate their capability profile.

GraphRouter \citep{Feng2024-GraphRouter} takes a different approach by modelling the relationships among tasks, queries, and LLMs through a heterogeneous graph structure. The method constructs a graph with task, query, and LLM nodes, where edges represent interactions between these entities. A Graph Neural Network (GNN) trained on historical performance and cost data performs edge prediction to forecast both the effectiveness and expense of routing a query to a specific LLM. This inductive learning framework enables GraphRouter to generalize to newly introduced LLMs without retraining, as it learns patterns from node characteristics rather than memorizing specific models. Experiments demonstrate substantial performance improvements while reducing computational overhead.

IRT-Router \citep{Song2025-IRT-Router} explicitly models two complementary properties for routing: the ability of each LLM, learned from its profile metadata, and the difficulty and discrimination of each query, extracted from its embedding. These are combined in an interactive layer to predict the probability that each candidate LLM will correctly answer the query. The predicted performance is then combined with each LLM's cost to produce a routing score, and the query is directed to the highest-scoring model. This formulation, inspired by Item Response Theory from educational psychometrics, provides interpretable routing decisions and handles unseen queries at deployment time through a semantic similarity warm-up mechanism.

A related line of work routes among a pool of already-trained LoRA adapters that share one base model, rather than among independently trained models themselves. LoraRetriever \citep{Zhao2024-LoraRetriever} retrieves relevant adapters by embedding each one from a handful of samples drawn from its own training data, using a retrieval encoder trained with a contrastive loss, then composes the top-ranked adapters for a query. LoRAuter \citep{Dhasade2026-LoRAuter} removes both requirements. It builds a small database of representative tasks from lightweight validation sets, and rather than matching a query to individual adapters directly, retrieves the tasks most similar to the query and composes the LoRA adapters trained for those tasks through an input-weighted combination. POLAR \citep{Li2026-POLAR} addresses a complementary edge-serving constraint, where only a small subset of a much larger adapter pool can stay resident in fast memory at once. It frames the joint decision of which adapters to keep resident and which adapter to route each request to as a two-timescale contextual bandit, and its POLAR+ variant proves a sublinear regret bound under added stationarity and cacheability assumptions on the workload.

While our survey focuses on multi-LLM routing, it is worth mentioning that difficulty estimation has also been explored in single-LLM contexts. Recent research found that LLMs frequently ``overthink'' simple queries and ``underthink'' complex ones \citep{Aggarwal2025-OptimalThinkingBench}, leading to inefficiencies. Moreover, sophisticated chain-of-thought reasoning primarily improves performance on complex maths and logic tasks, with limited gains for other tasks \citep{Wei2022-CoT-Prompting,Sprague2025-CoT-Math}.
On one hand, difficulty estimation can be calculated before sending the query to a reasoning LLM to decide whether to activate the thinking mode%
\footnote{Some approaches such as AdaptThink \citep{Zhang2025-AdaptThink} decide whether to activate the thinking mode from the beginning, while other approaches such as Continue-Thinking \citep{Zhang2025-ContinueThinking} first estimate the quality of the generated response without thinking to decide whether to continue generation with thinking.}
\citep{Zhang2025-AdaptThink,Zhang2025-ContinueThinking}, or to constrain the thinking token budget \citep{Shen2025-DAST-DifficultyAdaptive}. On the other hand, probing the LLM during the reasoning process can help identify whether to continue with further reasoning steps or to stop early \citep{Wu2025-ThoughtCalibration,Jiang2025-AdaptiveThinking}. This line of work can fall under adaptive computation approaches that aim to optimize resource usage within a single LLM.

%% file: sections/preference.tex
\section{Human Preference-aligned Routing}
\label{sec:preference}

In this paradigm, human preference data is used to train routers that generalize across domains and LLMs.
Within the design-space framework (cf.~Table~\ref{tab:framework}), these methods are characteristically pre-generation approaches that operate on query-level preference signals, with routing decisions computed via supervised classifiers or win-prediction models.
For example, RouteLLM \citep{Ong2025-RouteLLM} introduces a learning framework for training router models that dynamically select between strong and weak LLMs at inference time, optimizing the trade-off between response quality and cost. It formulates routing as a binary decision between strong (high-quality, high-cost) and weak (lower-quality, low-cost) LLMs, and employs a win prediction model to estimate the probability that the strong model will outperform the weak model for a given query. RouteLLM optimizes router parameters by maximizing the likelihood of observed preference data. Unlike Hybrid-LLM \citep{Ding2024-HybridLLM} and Arch-Router \citep{Tran2025-Arch-Router} that use solely synthetic preference data generated by an LLM, RouteLLM \citep{Ong2025-RouteLLM} leverages human preference labels from Chatbot Arena \citep{Chiang2024-ChatbotArena}. In addition, RouteLLM authors obtain pairwise comparison labels using an LLM judge. This data augmentation technique with synthetic preference labels substantially improves router performance. They experiment with four router architectures, namely similarity-weighted SW Ranking, matrix factorization, BERT, and causal LLM (based on Llama 3 8B). The matrix factorization router and causal LLM routers perform very competitively on  MT Bench \citep{Zheng2023-MTBench} when trained on the augmented dataset of both Chatbot Arena and LLM judge, with matrix factorization being the most efficient and cost-effective router.

Meta-Router \citep{Zhang2025-MetaRouter} instead addresses training-data quality: framing the choice between scarce gold-standard labels and abundant but biased preference comparisons, of the kind RouteLLM draws from Chatbot Arena, as a treatment-assignment problem, and correcting for this bias with a causal-inference objective.

\citet{Tsiourvas2025-CausalLLMRouting} address a different data limitation. Routers are typically trained on full-feedback data, where every candidate model has been evaluated on every query, which is expensive to collect and maintain as new models appear. They instead learn from observational data, where each historical query was answered by only the model actually deployed, using a doubly-robust causal estimator to correct for the resulting treatment-selection bias, and an end-to-end regret-minimization objective in place of the usual decouple-then-select pipeline.

\vspace{10pt}
\input{figures/ArchRouter}

Similarly, Arch-Router \citep{Tran2025-Arch-Router} aligns routing decisions with explicit user preferences. However, rather than optimizing purely for cost or performance, it allows users to define domain-action pairs. For instance, legal summarization queries can be routed to one model, while code generation requests will be sent to another model. The key insight is that routing policies are provided as part of the input context to a 1.5B parameter model, which means users can update their preferences without retraining. This is particularly valuable in production environments where routing requirements evolve frequently. Nevertheless, this flexibility comes with trade-offs, as a 1.5B model introduces additional computational overhead compared to lightweight classification-based routers, which may be suboptimal in latency-sensitive applications. Figure~\ref{fig:arch-router} illustrates the preference-aligned routing mechanism that Arch-Router has introduced.

Prompt-to-Leaderboard (P2L) \citep{Frick2025-Prompt-to-Leaderboard} extends traditional preference-based leaderboard evaluation by training an LLM to generate prompt-specific Bradley-Terry coefficients for model ranking. Unlike standard leaderboards that average performance across all tasks, P2L takes a user's prompt as input and outputs customized strength scores for each model, predicting which LLM will perform best for that particular query. The method enables task-specific evaluation, optimal cost-constrained routing, and personalized model selection based on user history.

GMTRouter \citep{Xie2025-GMTRouter} pursues personalization explicitly, modelling multi-turn user-LLM interaction histories as a heterogeneous graph over four node types, namely users, LLMs, queries, and responses, connected through per-turn nodes that aggregate each round of a conversation. User feedback, whether a rating, a pairwise ranking, or a reference response, is converted into a preference signal on this graph, and an inductive graph neural network learns to predict which LLM a given user will prefer for a new query. As the model is inductive rather than tied to a fixed set of users, it can personalize to a new user from only a handful of their past interactions, addressing the sparse, noisy feedback that limits routers relying solely on a single user's own data.

Eagle \citep{Zhao2024-Eagle} introduces a training-free approach to preference-aligned routing by using ELO ranking  systems \citep{Elo1967-Elo-Rating}. The method combines two complementary modules: Eagle-Global, which evaluates model overall capability using ELO ratings across the entire historical dataset, and Eagle-Local, which assesses specialized abilities through ELO rankings on similar past queries. By transforming sparse pairwise comparisons into comprehensive rankings, Eagle efficiently integrates user feedback without requiring model training.

While the aforementioned methods directly optimize against human or synthetic preference comparisons, some approaches rely on reward models trained on such data to provide supervisory signals.
Zooter \citep{Lu2024-Zooter} takes a reward-guided supervised learning approach, training a routing classifier using labels derived from the QwenRM reward model \citep{Bai2023-Qwen}, which scores candidate LLM responses on training queries.\footnote{QwenRM is typically used for RL training (e.g., GRPO) or SFT data construction through rejection sampling. Zooter uses it solely to generate routing labels.} The routing function outputs a categorical distribution over a pool of LLMs, representing the likelihood that each model can best handle the query. This knowledge distillation approach employs a BERT-style router (mDeBERTa-v3-base \citep{He2021-DeBERTaV3}). However, since training relies on a fixed set of LLMs, Zooter's adaptability to new or unseen models is limited.

%% file: figures/ArchRouter.tex
\begin{figure*}[htbp]
    \centering
    \includegraphics[width=1.0\linewidth]{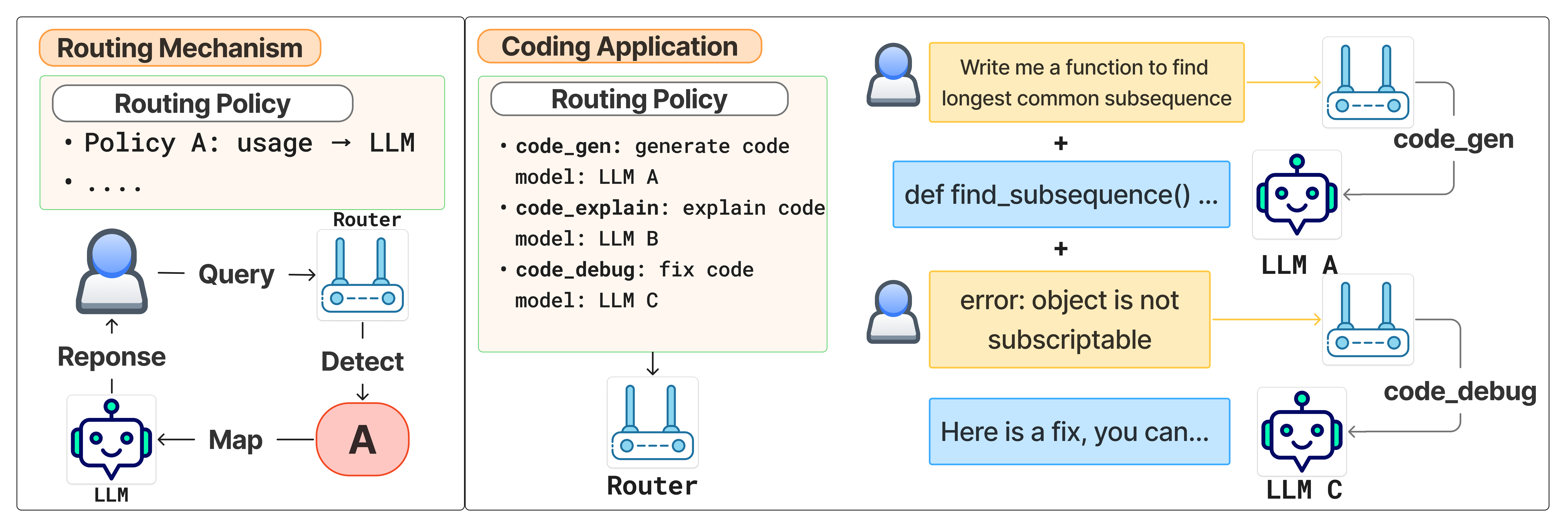}
    \caption{Arch-Router's Preference-Aligned Routing Mechanism. The routing policies and user conversation are provided to the router to select the appropriate policy and corresponding LLM. Example usage in coding is shown on the right.}
    \label{fig:arch-router}
\end{figure*}

%% file: sections/clustering.tex
\section{Clustering-based Routing}
\label{sec:clustering}

Clustering-based routing groups similar queries using unsupervised learning and assigns each cluster to the most suitable LLM. This approach balances performance and cost by routing different query types to their most cost-effective models, without requiring explicit task labels.

As illustrated in Figure~\ref{fig:uni-route}, UniRoute \citep{Jitkrittum2026-UniRoute} applies $K$-means clustering to an unlabelled training dataset to identify $K$ centroids, then partitions a validation set into these representative clusters. Each candidate LLM is evaluated on the validation data within each cluster, with scores adjusted by a predefined cost value for that model. At inference time, incoming queries are compared against the cluster centroids to determine which LLM should handle the request, with the accuracy-cost trade-off controlled through dynamic cost adjustment during routing.

Among the main practical advantages of this clustering approach is the ability to add new LLMs at inference time without retraining, as they are simply evaluated on existing clusters to obtain their performance profile. Moreover, the method can operate entirely without task labels, relying solely on query embeddings and cluster assignments. UniRoute demonstrates effective scaling, handling routing decisions across unseen LLMs.

\input{figures/UniRoute}

The Avengers \citep{Zhang2026-Avengers} introduces a general clustering-based routing recipe for combining smaller open-source models into a collective ensemble. Incoming queries are embedded and grouped into semantic clusters; each candidate model is scored per cluster during training; and at inference time each query is assigned to its nearest cluster and routed to the top-performing model(s), with repeated sampling and majority voting to aggregate outputs. This four-step paradigm---embed, cluster, rank, route---demonstrates that coordinated ensembles of smaller models can match or exceed the accuracy of proprietary models without modifying any individual model.
Avengers-Pro \citep{Zhang2025-Avengers-Pro} extends this paradigm by incorporating an explicit performance-efficiency trade-off. Each cluster assignment is governed by a scoring function that interpolates between correctness and inference cost via a single hyperparameter, tracing an accuracy-cost Pareto frontier across operating points.

Within the design-space framework (cf.~Table~\ref{tab:framework}), clustering-based methods are characteristically pre-generation approaches operating on query-level signals, with routing decisions made by unsupervised assignment to learned centroids. CRE-Router \citep{Moslem2026-CRE-Router} demonstrates that this paradigm can be integrated into multi-stage systems, proposing a two-stage cascaded solution where Stage~1 clusters incoming queries and assigns each cluster to its most cost-effective model, and Stage~2 adds a quality estimation cascade to escalate low-quality outputs to a stronger model, as discussed in detail in Section \ref{sec:cascades}.

%% file: figures/UniRoute.tex
\begin{figure*}[htbp]
    \centering
    \includegraphics[width=0.8\linewidth]{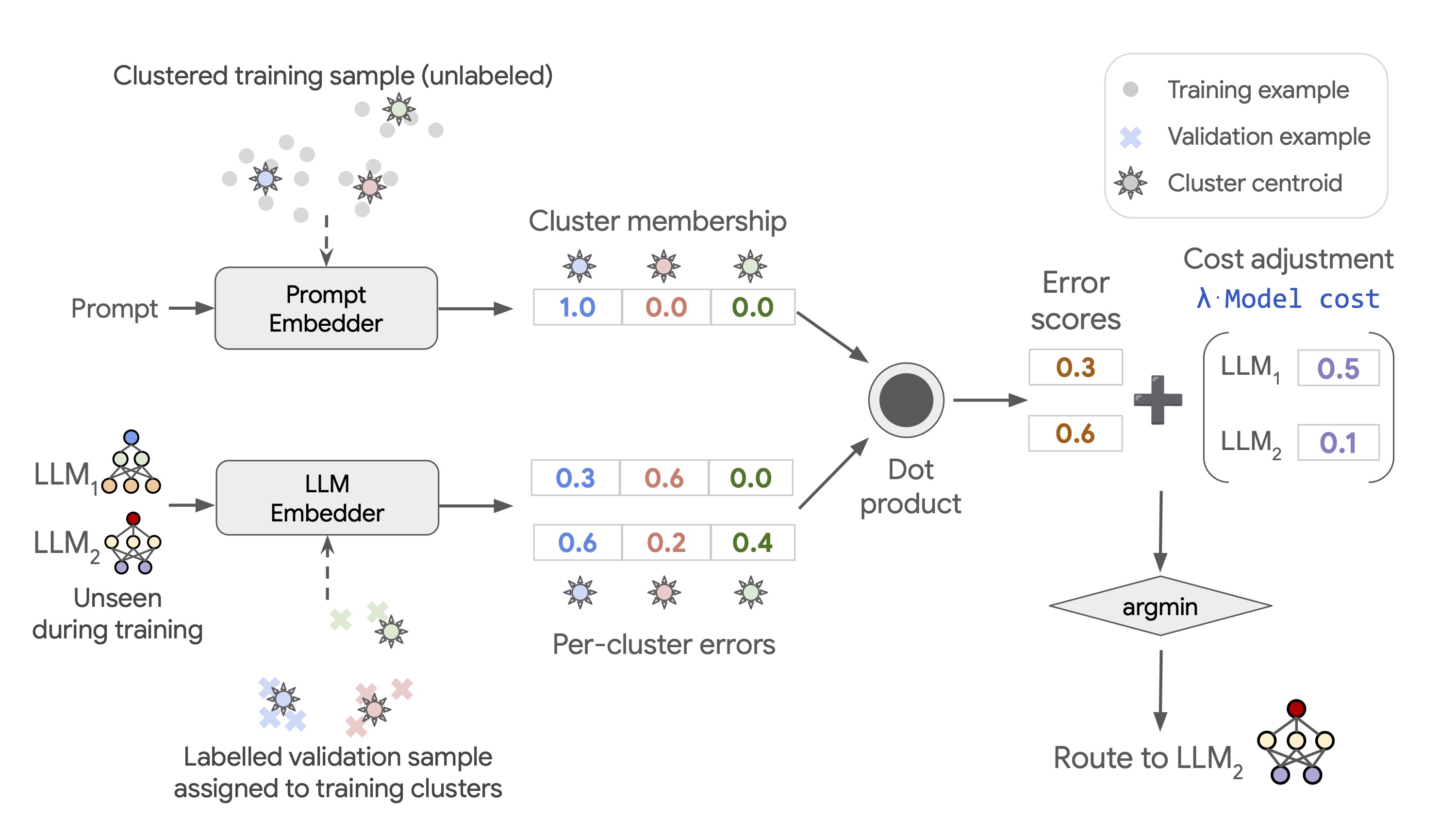}
    \caption{UniRoute cluster-based router. First, perform $K$-means on a training set to find K centroids, and then partition the validation set into K representative clusters. Each test-time LLM can then be represented as a $K$-dimensional feature vector of per-cluster errors. This yields an intuitive routing rule: for each test prompt, route to the LLM with the smallest cost-adjusted average error on the cluster the prompt belongs to. The prompt embedder may either be completely unsupervised (as shown in the figure), or fitted via supervised learning using labels from a set of training LLMs different from those seen during test time.}
    \label{fig:uni-route}
\end{figure*}

%% file: sections/rl.tex
\section{Reinforcement Learning Routing}
\label{sec:rl}

Reinforcement learning (RL) approaches formulate routing as decision-making under uncertainty.
Methods in this section include policy optimization algorithms that iteratively refine routing decisions through multi-step interactions and online bandit algorithms that balance exploration and exploitation with real-time feedback.
Within the design-space framework (cf.~Table~\ref{tab:framework}), policy optimization methods make multi-stage decisions drawing on query and model-level signals, while bandit methods are pre-generation approaches that adapt online through deployment feedback; both occupy the more complex end of the \textit{how} axis.

\subsection{Policy Optimization Methods}

Policy optimization approaches such as Router-R1 \citep{Zhang2025-Router-R1} and Route-and-Reason (R2-Reasoner) \citep{Shao2025-R2-Reasoner} optimize routing decisions through multi-step interactions. These RL approaches leverage the strengths of multiple LLMs to solve complex reasoning tasks while optimizing performance-cost trade-offs.
While Router-R1 addresses multi-round routing and aggregation, R2-Reasoner focuses on subtask-level decomposition.
Although the router in these methods is itself an LLM performing multi-step reasoning, all methods in this section dispatch queries to independently trained, separately deployed models, so the routing happens between model boundaries rather than within a shared architecture.

Router-R1 formulates routing as a sequential decision-making process. It alternates between performing internal reasoning (``think'' action) and assigning a specific model from a pool of available LLMs (``route'' action), iteratively refining answers. The router itself is an LLM that can leverage its reasoning ability to dynamically select and aggregate responses from multiple models. Router-R1 initiates a maximum of 4 routing steps per input query. The authors use Qwen2.5-3B-Instruct \citep{Yang2024-Qwen25} and LLaMA-3.2-3B-Instruct \citep{Dubey2024-Llama3} as base models. They are trained with the Proximal Policy Optimization (PPO) \citep{Schulman2017-PPO} reinforcement learning algorithm, on a training dataset of question–score pairs representing the quality of responses of each LLM. The authors argue that Router-R1 can generalize to new LLMs without retraining by conditioning on simple descriptors such as pricing and latency.



Similarly, R2-Reasoner decomposes complex reasoning tasks into simpler subtasks using a dedicated Task Decomposer. These subtasks are then allocated to the most suitable model by a Subtask Allocator. R2-Reasoner employs a staged training pipeline, combining supervised fine-tuning and reinforcement learning with Group Relative Policy Optimization (GRPO) \citep{Shao2024-DeepSeekMath}. This iterative approach refines the decision-making capabilities of the decomposer and allocator without requiring end-to-end gradient propagation. The authors argue that the framework achieves substantial cost reductions while maintaining competitive reasoning accuracy. It also generalizes well to unseen models, making it adaptable to dynamic real-world scenarios.

SCOPE (Scalable and Controllable Outcome Performance Estimator) \citep{Cao2026-SCOPE} takes a distinct approach by using GRPO to train a performance estimator rather than a routing policy. Given a query and a candidate model, SCOPE retrieves behavioural examples from a pre-computed model fingerprint and predicts both the model's correctness and token cost via chain-of-thought reasoning. The final routing decision is then made by a simple rule-based selection using a user-defined accuracy–cost trade-off, requiring no further learning. Unlike Router-R1 and R2-Reasoner, which use RL to directly optimize routing decisions, SCOPE's RL component is closer to reward-shaped prediction training than policy optimization. Furthermore, by characterizing models through behavioural fingerprints rather than fixed identifiers, SCOPE generalizes to unseen models without retraining.

xRouter \citep{Qian2025-xRouter} adopts a tool-calling formulation in which a router, trained end-to-end via reinforcement learning under an explicit cost-aware reward, can either answer directly or invoke external models, removing the need for hand-engineered escalation rules.

Although Router-R1, R2-Reasoner, and xRouter all aim to optimize cost, their multi-step or tool-calling formulations can still incur multiple model calls per query, introducing additional inference latency. These approaches are therefore best suited to scenarios where the resulting cost savings outweigh the added latency, such as routing to commercial LLM APIs with high per-call costs.

\subsection{Bandit-Based Methods}

Bandit-based methods frame routing as an online learning problem, balancing exploration of model capabilities with exploitation of known strengths while adapting to feedback over time.

MetaLLM \citep{Nguyen2025-MetaLLM} formulates routing as a multi-armed bandit problem, dynamically selecting the least expensive LLM likely to provide a correct answer for each query. Unlike Zooter's externally-trained reward model, MetaLLM's reward signal is computed directly from ground-truth correctness and cost, requiring no separate scoring model. Instead, it optimizes routing decisions based on an accuracy-cost trade-off, adapting to query difficulty and model performance over time.

MixLLM \citep{Wang2025-MixLLM} employs a contextual bandit framework with policy gradient methods to optimize query-LLM assignments. The system enhances query embeddings with domain-aware tags through unsupervised fine-tuning, enabling more accurate predictions of LLM-specific response quality and cost. A meta decision-maker selects assignments to balance quality, cost, and latency constraints. During online deployment, MixLLM updates its routing policy using binary user feedback through a dynamic feedback score mechanism, applying policy gradient optimization to maximize expected reward. This continual learning approach allows the system to adapt to evolving queries and changing LLM candidate sets.

PILOT (Preference-prior Informed LinUCB for Adaptive Routing) \citep{Panda2025-PILOT} extends contextual bandits by integrating offline human preference data with online evaluative feedback. The method reformulates routing as a contextual bandit problem where the system selects LLMs based on binary success/failure signals rather than full supervision. Building on the LinUCB algorithm \citep{Li2010-LinUCB} widely used in recommender systems, PILOT incorporates preference priors from sources like Chatbot Arena to accelerate learning. The authors additionally model cost constraints as an online multi-choice knapsack problem \citep{Zhou2008-Knapsack} to ensure resource-efficient routing under budget limits. This combination enables dynamic, personalized LLM selection that balances performance and cost.

GreenServ \citep{Ziller2026-GreenServ} extends the contextual bandit framework with an explicit focus on energy efficiency. Rather than optimizing for cost proxies such as API prices or token budgets, GreenServ measures GPU energy consumption directly during inference. It extracts a lightweight context vector per query, encoding task type, semantic cluster, and textual complexity, and then employs LinUCB \citep{Li2010-LinUCB} to learn routing policies online across a pool of 16 open-access LLMs. A tunable trade-off parameter $\lambda$ allows operators to interpolate between accuracy-first and energy-first policies.

LLM Routing with Duelling Feedback \citep{Chiang2025-Dueling-Feedback} addresses routing through a contextual duelling bandit formulation, learning from pairwise preference comparisons rather than absolute performance scores (cf.~Section~\ref{sec:preference}). The method introduces Category-Calibrated Fine-Tuning (CCFT), which constructs model embeddings by first fine-tuning text encoders with contrastive learning to cluster queries by category, then computing weighted combinations of category embeddings based on each LLM's domain expertise and cost profile. These learned embeddings enable the deployment of Feel-Good Thompson Sampling for Contextual Duelling Bandits (FGTS.CDB), a theoretically grounded Bayesian algorithm for online model selection. Evaluations on RouterBench and MixInstruct demonstrate that the approach achieves lower cumulative regret and faster convergence than baselines using general-purpose embeddings, while maintaining robust performance-cost balance.

Similarly, Time-Increasing UCB (TI-UCB) \citep{Xia2024-TI-UCB} addresses online model selection through a time-increasing bandit algorithm that accounts for the increasing-then-converging performance trend observed during iterative model fine-tuning, a common pattern as LLMs are progressively improved. Unlike traditional bandit methods that assume stationary reward distributions, TI-UCB incorporates a change detection mechanism to identify convergence points, enabling more efficient exploration-exploitation balance. The approach achieves logarithmic regret bounds and demonstrates practical effectiveness in both classification model selection and online LLM selection scenarios.

\citet{Bae2026-CQB-MNL} extend the bandit framing to jointly address routing and scheduling: rather than assuming queries are served in arrival order, their contextual queueing bandit decides both which pending query to serve next and which LLM to route it to. The key idea is to learn from implicit feedback inferred from user retrials, treating a user re-submitting a query as a signal of dissatisfaction rather than requiring explicit ratings or preference comparisons. Their algorithm, anytime CQB, combines Thompson sampling with a decaying schedule of forced exploration, extending closely related queueing-bandit theory to a multinomial-choice setting and to an anytime regime that needs no advance knowledge of the serving horizon or traffic load, both required for continuous LLM-serving deployment.

%% file: sections/uncertainty.tex
\section{Uncertainty-based Routing}
\label{sec:uncertainty}

Researchers have been investigating whether there is a reliable way to estimate if a model is confident about its own response, where such confidence must be aligned with actual correctness. This notion has several applications in routing, as effective uncertainty quantification enables systems to identify when to escalate queries to more capable models.
Uncertainty-based methods are post-generation approaches that use response-level signals, i.e. confidence scores or token probabilities, to trigger escalation decisions.

\subsection{Probing and Probability-based Uncertainty Estimation}

In their work, \citet{Chuang2025-ConfidentSeekStronger} benchmark eight uncertainty quantification methods for routing from SLMs to LLMs on edge devices. They observe that probe-based methods (trained classifiers) and perplexity-based methods outperform verbalization approaches (self-reported confidence). They demonstrate that SLMs match LLM performance on high-confidence (top 20\%) queries. These results are in line with the experiments by \citet{Mahaut2024-FactualConfidence} who demonstrate that trained hidden-state probes provide more reliable confidence estimates than alternatives, even though at the expense of requiring access to weights and supervision data.

CP-Router \citep{Su2025-CP-Router} applies Conformal Prediction to route between standard LLMs and Large Reasoning Models (LRMs) like DeepSeek-R1, which often generate verbose reasoning for simple queries. For multi-choice question answering, the method extracts logits for answer options, applies softmax to obtain a probability distribution, and computes uncertainty scores as 1 minus the probability for each option. It then builds a prediction set based on these scores against a calibration-derived threshold. Queries with a single plausible option (high confidence) are handled by the standard LLM, while those with multiple plausible options (high uncertainty) are routed to the more capable but costly LRM.

\citet{Zhang2025-LLMJudge-UncertaintyRouting} train a predictive router on preference labels derived from semantic entropy. For each query, they sample multiple responses from a strong and a weak model, cluster them by bidirectional entailment, and compare the resulting entropy between the two models to determine which one should win. The trained router itself is a pre-generation classifier over query embeddings, following the same architecture family as RouteLLM (similarity-weighted ranking, matrix factorization, MLP, and k-nearest neighbours), so no response is generated before the routing decision at inference time. The authors separately introduce LLM-as-a-Judge as a protocol for evaluating response quality across routing methods, but it plays no role in the router itself.

\subsection{Self-Reported Confidence Estimation}

Instead of confidence calculations, researchers employ self-verification or self-reporting, where an LLM is asked to report its own confidence. \citet{Chuang2025-ConfidentSeekStronger} refer to these approaches as ``verbalization'' and classify them as one-step verbalization, where an LLM is prompted to output both the answer and numeric confidence in one step, and two-step verbalization, where the confidence score is obtained in a separate, follow-up query after the model has provided an answer \citep{Tian2023-AskCalibration}. 
However, their experiments indicate that both verbalization techniques consistently exhibit low alignment between reported uncertainty and prediction correctness.

Despite these limitations, recent work demonstrates that self-verification can be effective in cascading systems when combined with other mechanisms. AutoMix \citep{Aggarwal2024-AutoMix} uses few-shot in-context learning for self-verification, while Self-REF \citep{Chuang2025-Self-REF} employs lightweight fine-tuning to improve confidence calibration. These methods are discussed in detail in Section~\ref{sec:cascades} on cascading approaches, as they rely on self-verification as part of multi-stage routing pipelines rather than as standalone uncertainty quantification.

%% file: sections/cascades.tex
\section{Cascades}
\label{sec:cascades}

While routing preselects an appropriate LLM for a  query, cascading sequentially queries a pool of LLMs until a reliable response is obtained. Cascading strengthens the routing process by allowing multiple LLMs to be involved in answering a single query. In a typical cascading setup, a smaller and cheaper LLM is first queried to generate an initial response. Based on the quality of this response, the system decides whether to accept it or escalate the query to a larger and more capable LLM for further refinement or regeneration. This cascading system design can be particularly useful when dealing with queries of varying difficulty levels, as it allows the system to adaptively allocate resources based on the complexity of the task at hand.
Within the design-space framework (cf.~Table~\ref{tab:framework}), cascading systems are inherently multi-stage, combining query-level and response-level signals across successive models, and often compose multiple mechanisms within a single pipeline.

\input{figures/AutoMix}

FrugalGPT \citep{Chen2024-FrugalGPT} mixes routing and cascading through two components: an LLM router and a quality-scoring function. The router selects a fixed, query-agnostic list of LLMs to try in sequence, learned once offline over the training distribution rather than adapted per query. Given a user query $q$, each LLM in the list is invoked in turn, and the quality-scoring function, based on DistilBERT \citep{Sanh2019-DistilBERT}, takes the query, the response, and the selected LLM as input to generate a reliability score. If this score clears a per-step threshold, learned jointly with the router's list, the answer is returned; otherwise the next LLM in the list is queried. Similarly, \citet{Dekoninck2025-CascadeRouting} propose cascade routing, a unified framework that integrates routing and cascading into a single system. Unlike pure routing, which selects a single model per query, or cascading, which runs models sequentially from smallest to largest, cascade routing iteratively selects the best model at each step, allowing it to skip or reorder models dynamically. The authors identify quality estimation as the critical factor for model selection success and show that cascade routing outperforms individual approaches.

CRE-Router \citep{Moslem2026-CRE-Router} illustrates how clustering and cascading can be composed into a unified pipeline in two pre-generation and post-generation stages. Stage~1 assigns incoming queries to semantic clusters and routes each cluster to its most cost-effective model using a scoring function that trades off error rate against a chosen cost metric, with the cost budget controlled by an interpretable hyperparameter tuned offline on task-correctness labels (cf.~Section~\ref{sec:clustering}). Stage~2 then applies a quality estimation cascade. A ModernBERT-based binary classifier, trained on the same task-correctness labels as Stage~1, inspects each efficient-model output together with its response length and escalates queries it labels low-quality to a stronger model. Outputs already produced by a strong model in Stage~1 bypass this classifier entirely, so the cascade only adds inspection cost where an efficient model was used in the first place. Collective pre-generation routing at the cluster level avoids running efficient models on query groups they are unlikely to handle correctly, while the post-generation quality estimation cascade recovers accuracy on individual queries where the efficient model falls short. Together, the two stages span the design space from pre-generation cluster-level routing to post-generation response evaluation, with the framework adapting to model pool changes through automatic Pareto analysis.
\citet{Moslem2026-CRE-Router} show that the choice of cost metric itself matters. Throughout, cost is measured in latency rather than monetary price. Decoding speed (TPOT) divides out a response's length, rating a terse model and a verbose one as equally cheap per token even though the verbose model costs far more per request, while end-to-end latency (E2EL) penalizes that verbosity directly. Fitting the same routing framework separately under each metric yields materially different routing decisions, and no single proxy such as thinking mode reliably predicts which models this affects. They address this by blending the two costs into a single convex combination within the Stage~1 score. At equal weighting, this blended cost selects at least one routing that neither pure metric would choose on its own.
SATER \citep{Shen2025-SATER} pursues a similar goal of unifying pre-generation routing and cascade routing, but instead of relying on an external quality estimator or classifier as in FrugalGPT and CRE-Router, it fine-tunes the models in the pool directly, using shortest-response preference optimization together with a confidence-aware rejection mechanism built into the model itself. The authors suggest that their approach reduces redundant, overly verbose outputs and cuts computational cost and cascade latency while preserving accuracy.

Instead of relying on an external model for quality estimation, some researchers have proposed ``self-verification'' as an efficient approach for model cascading \citep{Aggarwal2024-AutoMix, Chuang2025-Self-REF}. In this paradigm, an efficient LLM learns to estimate the quality of its own response. The self-verification score is then used to decide how to proceed. For instance, AutoMix \citep{Aggarwal2024-AutoMix} uses a smaller model to generate an initial answer and self-verify its response before potentially routing the query to a larger model. The approach relies on few-shot prompting without fine-tuning, making it suitable for black-box models. While previous work finds self-verification unreliable to repair the model's output \citep{Huang2024-LLM-SelfCorrect} or estimate its confidence \citep{Mahaut2024-FactualConfidence,Chuang2025-ConfidentSeekStronger}, \mbox{AutoMix} demonstrates that few-shot self-verification provides a valuable signal that can be used for routing to an appropriate model. The final step involves using a router to decide whether to escalate the query to a larger model if it is a hard query, or output the current answer to the user if it is easy or unsolvable.%
\footnote{The idea of blocking ``unsolvable'' queries from being routed to the larger model as an efficiency booster has been investigated in detail in recent work such as Firewall Routing \citep{Peng2025-FirewallRoutingLLMs}.}
The router is modelled as a Partially Observable Markov Decision Process (POMDP) \citep{Astrom1965-POMDP,Wu2021-AdaOpsPOMDP}, which relies only on self-verification confidence as input.
Figure~\ref{fig:auto-mix} illustrates the AutoMix approach.

In the domain of code generation, \citet{Chen2025-ModelCascadingCode} apply a related self-generated escalation signal. Each candidate model in the cascade produces both code solutions and executable test cases for a query, and solutions are scored by how many of the model's own generated tests they pass, extending CodeT's within-model ranking scheme to a cross-model escalation decision. When the best-scoring solution clears a threshold calibrated in advance on a held-out set of cost-accuracy trade-off points, it is returned; otherwise the query escalates to the next larger model in the family. Unlike FrugalGPT's trained quality estimator, this signal comes directly from test execution rather than a learned scorer, and the pipeline remains entirely black-box.

While AutoMix relies solely on few-shot in-context learning without any fine-tuning, Self-REF \citep{Chuang2025-Self-REF} introduces a lightweight fine-tuning strategy that enables LLMs to express confidence in their predictions using special confidence tokens. This confidence signal is then used for downstream routing, where queries with low confidence are routed to stronger (but more costly) LLMs, or rejection where the system should fall back on a safe default behaviour. Self-REF augments training data with confidence tokens based on correctness, fine-tunes the LLM to generate these tokens, and extracts a confidence score for each prediction.\footnote{To obtain a continuous confidence score after training, the authors compute the probability of the confident token \textit{<CN>}, normalized over the sum of the probabilities of the unconfident \textit{<UN>} and confident \textit{<CN>} tokens.}
This confidence-aware fine-tuning provides a practical mechanism for both routing and rejection in LLM cascades.

Beyond single-model confidence estimation, ensemble approaches take a complementary direction by querying multiple models simultaneously. LLM-Blender \citep{Jiang2023-LLM-Blender} uses an ensemble framework that queries multiple LLMs during inference, selects the best responses, and then merges them into one response. The framework consists of two main modules: Pair Ranker, which uses pairwise comparisons to rank candidate outputs from multiple LLMs, and Gen Fuser, which fuses the top-ranked outputs to generate an improved response. First, the Pair Ranker employs a cross-attention encoder to jointly encode the input and candidate pairs, achieving high correlation with human and ChatGPT-based rankings. Then, the Gen Fuser further enhances output quality by synthesizing the best aspects of top candidates. The approach is evaluated on the MixInstruct benchmark, demonstrating significant improvements over individual LLMs and baseline ensembling methods.

At a finer granularity than query-level cascading, R2R \citep{Fu2025-R2R} routes individual tokens between a small and large model, escalating only the small fraction of tokens where their reasoning paths genuinely diverge.\footnote{Unlike speculative decoding \citep{Leviathan2023-SpeculativeDecoding,Chen2023-SpeculativeSampling}, where a draft model's tokens are verified and accepted/rejected by the target model to exactly preserve its output distribution, R2R trains a router to decide upfront which model generates each token, with no verification step and no distributional guarantee.}

While several cascading approaches restart generation from scratch when a query is escalated to a more capable model, others simply refine the response already generated by the smaller model. In the area of machine translation (MT), cascaded system designs have been extensively studied and deployed in production. Work on automatic post-editing (APE) formulates translation as a two-stage process, in which a base MT system produces an initial hypothesis that is selectively refined by a more capable model when necessary. Usually, quality estimation (QE) models are used to score initial translations, and the routing system decides whether to accept a translation or to escalate to an APE model depending on predefined criteria or thresholds \citep{Specia2018-MT-QE,Hendy2023-LLM-MT,Moslem2023-AdaptiveMT,Zerva2024-WMT-QE}. This cascading approach has also been extended to terminology-constrained APE, instructing an LLM to integrate the required terms into the translations \citep{Moslem2023-Terminology}. Skipping post-editing for translations that already include pre-approved terms can reduce the overall post-editing load. Such an efficient workflow can save resources, and minimise latency at inference time. Similarly, there are potential advantages of employing an LLM for post-editing rather than for direct translation. Instead of solely relying on the translation quality of the LLM, quality estimation can be performed to select the best MT model in general or for the current source text segment. Ultimately, only segments that do not meet quality criteria are then passed to the LLM for post-editing. Furthermore, QE-based deferral has also been applied directly to routing between small and large translation models, bypassing post-editing entirely. \citet{Farinhas2025-TranslateSmart} use QE metrics as deferral rules in a cascaded system, invoking the larger model only for a fraction of segments. These cascading approaches demonstrate that quality estimation can effectively guide escalation decisions in hybrid model deployment, reducing large model calls while maintaining quality.

%% file: figures/AutoMix.tex
\begin{figure*}[ht]
    \centering
    \includegraphics[width=0.8\linewidth]{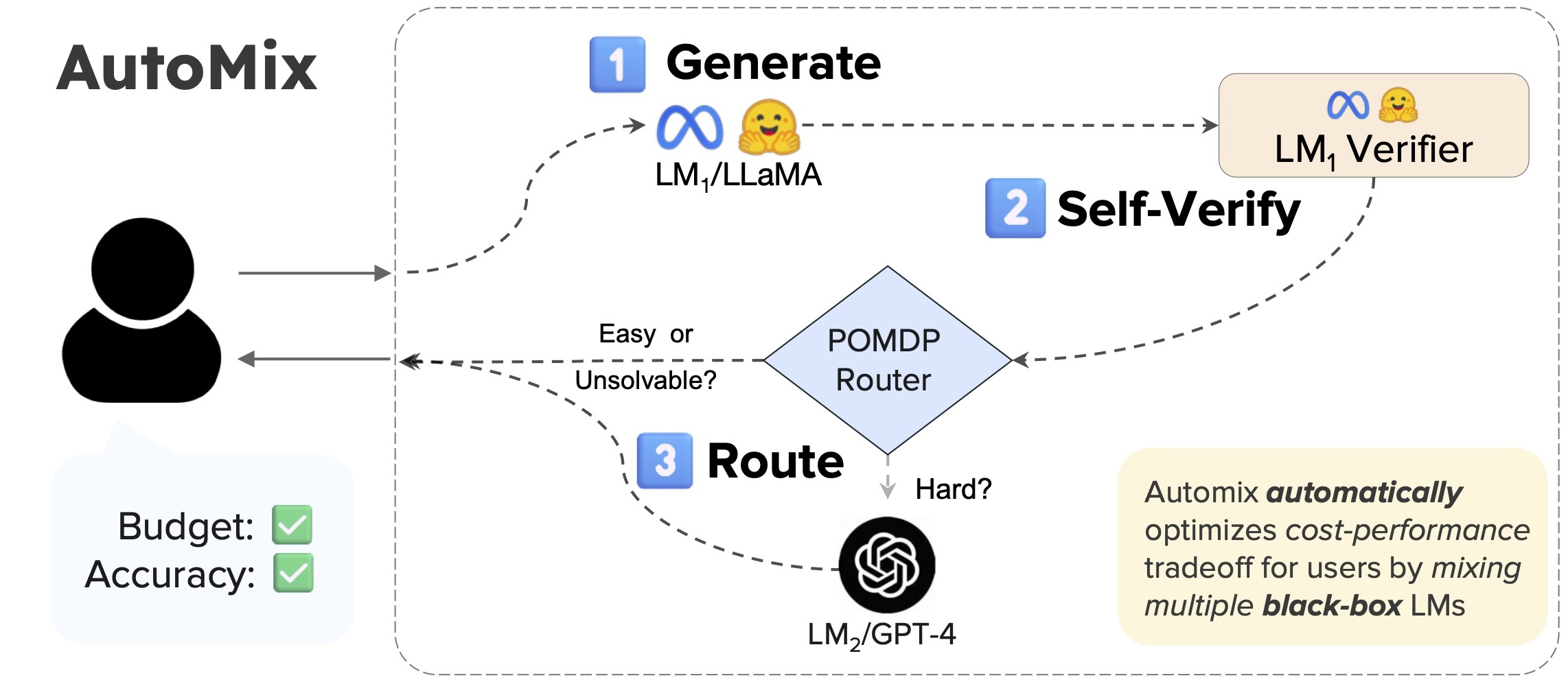}
    \caption{AutoMix example for two-model setup. Instead of relying only on a small model (SLM) with low performance or a large model (LLM) with high cost, AutoMix automatically mixes multiple black-box language models, based on user desired cost-quality tradeoff. AutoMix works in a 3-step process: (i) generation by a small model (LM1), (ii) self-verification of the generated answer, (iii) using confidence assessments from self-verification to do appropriate routing to a larger model (LM2). For N-model setup, the process is repeated till the final answer is reported.}
    \label{fig:auto-mix}
\end{figure*}

%% file: sections/multimodal.tex
\section{Multimodal Model Routing}
\label{sec:multimodal}

Research on model routing has been focused on text-based LLMs. Many of these approaches can be extended to other modalities such as speech, image, and video. However, representations of these modalities are different from text, which poses new challenges.

Model-Spider \citep{Zhang2023-Model-Spider} is a model selection and ranking method designed to efficiently choose suitable pre-trained models from a large model pool, including both visual models and LLMs, for a downstream task. Unlike the multi-LLM routing mechanisms we covered so far, Model-Spider does not dynamically dispatch queries to different models at inference time. It mainly evaluates candidate models from the available pool and ranks them according to predicted task fitness.

ReLope \citep{Zeng2026-ReLope} extends probe-based routing to multimodal LLMs (MLLMs), where visual inputs introduce new challenges for efficient deployment. The authors highlight the limitation that hidden-state probes effective in text-only LLMs degrade substantially when visual inputs are introduced, as visual tokens weaken the separability of correctness signals in hidden states. To address this, they  propose two complementary techniques. The Attention Probe aggregates token-level hidden states from the preceding layer using attention scores to recover distributed correctness signals. ReLope (KL-Regularized LoRA Probe) inserts a lightweight LoRA adapter and applies a KL regularizer to learn routing-aware representations that suppress task-irrelevant visual noise. 
Experiments across five multimodal benchmarks demonstrate improvements over existing probe-based baselines in hybrid MLLM systems.

MMR-Bench \citep{Haoxuan2026-MMR-Bench} is a modality-aware evaluation framework. It covers a wide range of vision-language tasks, including OCR, general VQA, and multimodal math reasoning. The benchmark includes strong single-model baselines, oracle upper bounds, and representative routing policies, enabling systematic comparison of routing strategies under fixed candidate sets and cost models.

VL-RouterBench \citep{Huang2026-VL-RouterBench} is another considerably larger-scale benchmark for VLM routing, covering 14 datasets across three task groups, 30,540 samples, and 17 candidate models (15 open-source and 2 API-based), yielding over half a million sample-model pairs. The benchmark evaluates ten routing methods and baselines, including VLM adaptations of RouterDC \citep{Chen2024-RouterDC} (cf.~Section~\ref{sec:difficulty}) and Zooter \citep{Lu2024-Zooter} (cf.~Section~\ref{sec:preference}), alongside routing architectures specific to the multimodal setting. Its main finding echoes the text-only routing literature: learned routers deliver a significant routability gain over any single model at comparable or lower cost, yet even the best-performing router, an adapted RouterDC, still shows a clear gap to the oracle upper bound, which the authors attribute to the difficulty of jointly modelling fine-grained visual cues and textual structure.

Despite these benchmarking efforts, multimodal routing remains underexplored compared to text-only settings. Key challenges include developing unified representations across modalities, handling queries that require multiple modalities simultaneously, and adapting routing strategies to modality-specific characteristics. Future work should explore how difficulty estimation, uncertainty quantification, and preference alignment can be extended from text-only to multimodal settings.

%% file: sections/evaluation.tex
\section{Evaluation}
\label{sec:evaluation}

In order to optimize routing efficiency, it is imperative to establish the criteria by which routing systems are evaluated. This section reviews commonly used benchmarks and metrics for assessing routing quality, efficiency, and cost, examines how these are applied in practice, and discusses common routing failure modes.

\subsection{Routing Benchmarks}
\label{subsec:benchmarks}

Several specialized benchmark datasets have been developed to evaluate LLM routing systems systematically.

RouterBench \citep{Hu2024-RouterBench} has been widely adopted for evaluating routing methods. It provides over 405k precomputed inference outputs from eleven diverse LLMs across seven tasks (MMLU, MT-Bench, MBPP, HellaSwag, WinoGrande, GSM8K, ARC). The dataset includes detailed performance and cost metadata, enabling systematic analysis of routing strategies under realistic constraints.

RouterEval \citep{Huang2025-RouterEval} is a large-scale open-source benchmark for systematic evaluation of LLM routing methods. The benchmark comprises over 200 million performance records from more than 8,500 LLMs across 12 widely used benchmarks, including ARC, HellaSwag, MMLU, TruthfulQA, GSM8K, and MMLU-PRO. RouterEval frames routing as an m-way classification problem with varying difficulty levels (m = 3, 5, 10, 100, 1000) and supports evaluation across all-strong, all-weak, and mixed model groups. Key findings demonstrate that capable routers can achieve performance surpassing the best single model through model complementarity.

MixInstruct \citep{Jiang2023-LLM-Blender} is an instruction-following benchmark designed to evaluate routing and ensemble methods using preference-based supervision. It aggregates 110k examples from multiple instruction datasets and provides oracle pairwise preferences obtained via LLM-based comparisons.

LLMRouterBench \citep{Li2026-LLMRouterBench} comprises over 400K instances from 21 datasets and 33 models. Moreover, it provides comprehensive metrics for both performance-oriented routing and performance-cost trade-off routing, and integrates 10 routing baselines. The LLMRouterBench framework integrates collector, evaluator, and adaptor components for standardized evaluation of LLM routing methods.

In addition to the aforementioned benchmarks, several further evaluation efforts target specific limitations in existing practice. RouterArena \citep{Lu2026-RouterArena} introduces an open, leaderboard-style platform for comparing routers across a broad range of knowledge domains and difficulty levels. The DSC benchmark \citep{Kassem2026-DSC} extends evaluation beyond task accuracy to privacy and safety, showing that preference-based routers can make category-driven decisions that inadvertently route jailbreaking attempts to weaker, less-safe models. \citet{Yuan2025-WhoRoutes} build on RouterBench and EmbedLLM-style query embeddings (cf.~Section~\ref{sec:difficulty}) to argue that existing router benchmarks suffer from limited task diversity, imbalanced model pools, and oversimplified evaluation methodologies, and propose a broader evaluation covering 68 task categories and 85 candidate models.

Router robustness under adversarial manipulation has also emerged as a concern. \citet{Shafran2025-ReroutingLLMRouters} show that a short adversarial token sequence prepended to any query can change a router's classification so that the query is sent to the more expensive model instead of the cheaper one, without altering the query's apparent content or degrading response quality, and that perplexity-based filtering fails to catch these sequences. They frame this as a risk because it lets an adversary inflate inference costs or obtain higher-quality responses than its routing policy intends to provide, undermining the cost or quality guarantees routing is meant to enforce. \citet{Tang2026-R2A} pursue the same purely black-box setting with a more principled method, optimizing an adversarial suffix against a surrogate router trained to mimic the target, again redirecting queries toward the more expensive model without access to the target router's internals. \citet{Yang2026-AmongUs} study a complementary threat. Rather than an external attacker manipulating the query, one of the models in the pool itself is malicious, engineered via prompting, activation steering, fine-tuning, or reinforcement learning to produce wrong, unsafe, or misleading outputs. They show this degrades performance across routing and three other multi-model collaboration paradigms, and find that an external supervisor model that screens and can disable suspicious pool members recovers most of the lost performance.


\subsection{Evaluation Metrics}
\label{subsec:metrics}

Evaluating routing systems requires metrics that capture both the quality of routing decisions and their computational efficiency. We organize these metrics into performance, efficiency, and cost metrics.

\subsubsection{Performance and Quality Metrics}

The fundamental goal of routing is to select models that produce high-quality responses. \textbf{Routing accuracy} measures the percentage of queries routed to the optimal model. In addition, a confusion matrix identifies false positives and false negatives, which helps calculate the overall performance of the routing system on the current task.

While routing accuracy measures whether the router has selected the ``optimal'' model for each query, task performance determines whether the selected model has produced a correct or high-quality response. \textbf{Task performance} evaluates response quality using domain-specific metrics such as accuracy on multiple-choice questions, exact match for factual queries, pass@k \citep{Kulal2019-Pass@K,Chen2021-EvaluateCode-Pass@K} for code generation, and chrF \citep{Popovic2017-chrF++} or COMET \citep{Rei2020-COMET} for translation. Moreover, LLM-as-a-Judge evaluations are increasingly adopted, where a strong LLM ranks the quality of responses from different models. The overall performance of a routing system is the aggregate task performance across all routed queries, effectively treating the router as a meta-system that combines multiple models.

For preference-based routing methods (cf.~Section \ref{sec:preference}), \textbf{win rate} measures how often the routed model's response is preferred over a baseline in pairwise comparisons, whether through human judgements or LLM-as-a-Judge evaluations. 
\textbf{Area Under Curve (AUC)} is used to summarise routing performance across different operating points, such as performance across varying cost budgets or deferral thresholds, providing a single metric for comparing routers independent of specific cost constraints.
RouterBench \citep{Hu2024-RouterBench} introduces \textbf{AIQ} (Average Improvement in Quality), which averages the accuracy-cost trade-off curve over a fixed cost range. RouteLLM \citep{Ong2025-RouteLLM} introduces \textbf{APGR} (Average Performance Gap Recovered), which integrates the fraction of the weak-to-strong performance gap recovered across cost budgets, and a complementary threshold metric, \textbf{CPT($x\%$)} (Call-Performance Threshold), the minimum percentage of queries that must be routed to the strong model to recover $x\%$ of that gap; lower values indicate a more cost-effective router.
\citet{Wu2026-RouterXBench} argue that these curve-based aggregate metrics conflate two distinct quantities: how well the router discriminates between queries the small model can and cannot handle, and how well a chosen deployment threshold matches a specific cost-quality requirement, so a high AUC or AIQ score can reflect the large model's inherent strength rather than the router's skill. They propose disentangling these with a threshold-independent discrimination measure, evaluation metrics computed separately for low-cost, balanced, and high-accuracy deployment regimes, and systematic scoring across in-distribution and out-of-distribution task pairs, arguing that a single blended curve score obscures both how a router behaves in the scenario an operator cares about and whether that behaviour holds outside its training distribution.

\subsubsection{Efficiency and Cost Metrics}

While routing aims to reduce computational costs, it introduces additional system overhead that must be carefully controlled. Obviously, adding larger models to the LLM routing pool contributes to this overhead. Additionally, the overhead resulting from the routing process itself must be negligible.

\textbf{Latency} refers to end-to-end latency (E2EL), the total time between sending a request and receiving a complete response. This is commonly decomposed into the time to first token metric (TTFT) during prefilling and the time per output token metric (TPOT) during decoding \citep{Huyen2025-AIEngineering}.

\textbf{Throughput} captures system capacity, and it is measured in tokens per second (TPS) or queries per second (QPS). There is an inherent trade-off between latency and throughput, where serving requests quickly (low latency) often conflicts with serving many requests efficiently (high throughput). \textbf{Goodput} measures throughput that meets predefined constraints.%
\footnote{Efficient inference frameworks such as vLLM \citep{Kwon2023-vLLM} offer server versions that can report several standard metrics such as TTFT, TPOT, throughput, and goodput under constraints such as concurrency and maximum token budget.}

\textbf{Cost} represents API charges, compute resources, and token consumption. Typically, cost can be reported relative to strong baselines (e.g., ``97\% of GPT-4's quality at 24\% of the cost'') or as a simple monetary value. To jointly evaluate quality and efficiency, quality-cost trade-offs are visualized through \textbf{Pareto frontiers} that plot achievable combinations of performance and cost, where effective routers should dominate any single model.

Beyond direct costs, routing systems can reduce environmental footprint by favouring smaller models. \textbf{Energy consumption} can be reported per token, per query, or as total inference energy, with lower consumption indicating improved efficiency. \textbf{Carbon footprint} measures $CO_2$ emissions, estimated from energy consumption and grid carbon intensity. These environmental metrics are increasingly adopted as environmental sustainability becomes central to efficient large-scale deployment.

\subsection{Evaluation of Routing Systems in Practice}

Recent work on multi-LLM routing and cascades has grown rapidly. As part of this effort, researchers have developed specialized routing benchmarks (cf.~Section~\ref{subsec:benchmarks}) or adopted widely used general-purpose LLM benchmarks. Nevertheless, systematic evaluation of routing systems still lacks maturity, and cross-examination of routing approaches remains limited.
Tables~\ref{tab:routerbench}, \ref{tab:mtbench}, and \ref{tab:mixinstruct} demonstrate approaches that rely on RouterBench \citep{Hu2024-RouterBench}, MT-Bench \citep{Zheng2023-MTBench}, and MixInstruct \citep{Jiang2023-LLM-Blender}, respectively. Across the three tables, every method outperforms at least one baseline under its own evaluation protocol, but no metric, model-pool size, or cost basis is shared across rows. On RouterBench, AUC (Cascade Routing), AIQ (GreenServ), accuracy-at-a-cost-point (IRT-Router, MixLLM, PILOT), and area-under-deferral-curve (UniRoute) are different ways of scoring ``performance'' over pools ranging from an unspecified size up to all 11 RouterBench models. On MT-Bench, the four methods express their gains in different units, namely a quality-drop percentage at a chosen cost-reduction level (BEST-Route), calls-to-target-accuracy as a \% of queries (RouteLLM) or converted to USD (Confidence-Driven Router, which adopted RouteLLM's own metric to compare against it directly), and a raw judge score (Zooter). Hence, drawing from the same dataset does not imply a shared evaluation protocol.

\input{tables/evaluation}

\subsection{Routing Failure Modes}

Understanding how and why routing systems fail is as important as demonstrating where they succeed. We highlight four such modes, namely distribution shift in the query stream, router staleness as the underlying model pool changes, the asymmetric cost of router errors, and a systematic bias toward over-selecting the most capable model as budget increases.

\paragraph{Distribution shift.} Most routers are trained on a fixed sample of queries and model responses, and their accuracy can degrade when deployed queries diverge from that training distribution. Few surveyed methods are explicitly evaluated under this condition. BEST-Route \citep{Ding2025-BEST-Route} reports an out-of-distribution evaluation, and IRT-Router \citep{Song2025-IRT-Router} addresses unseen queries directly through a semantic similarity warm-up mechanism. \citet{Dann2025-PrincipledModelRouting} address a more precise notion of shift, where the training data spans several known domains and their relative proportions at deployment are unknown and may vary. They train the router via a minimax game that adversarially reweights domain importance during training, and prove a worst-case regret bound, showing that for any mixture of the training domains, the router's performance stays close to what a perfectly mixture-aware router would achieve. This absence of distribution-shift testing is a real gap in how routing systems are currently evaluated.

\paragraph{Router staleness.} A router trained on a fixed model pool becomes stale as that pool changes, whether models are added, removed, or updated by their providers. Several methods are explicitly designed to avoid this, namely GraphRouter \citep{Feng2024-GraphRouter}, ICL-Router \citep{Wang2026-ICL-Router}, UniRoute \citep{Jitkrittum2026-UniRoute}, and CRE-Router \citep{Moslem2026-CRE-Router}; all generalize to new models without retraining the router itself, either through inductive model representations or by automatically re-running a selection procedure (e.g., CRE-Router's Pareto analysis) when the pool changes. By contrast, methods built around a fixed model pool, such as Zooter's reward-distilled classifier, whose limited adaptability to unseen models the original discussion already notes (cf.~Section~\ref{sec:preference}), are more exposed to this failure mode. TI-UCB \citep{Xia2024-TI-UCB} addresses a related but distinct form of staleness, non-stationary reward distributions as a single model's performance shifts during iterative fine-tuning, through an explicit change-detection mechanism. Feedback-adaptive methods more broadly carry a related risk beyond technical staleness (cf.~Section~\ref{sec:ethics}).

\paragraph{Cost of router errors.} Router mistakes are not symmetric in consequence. Sending a query beyond a weak model's ability risks an incorrect response, while sending a simple query to a strong model only wastes cost and latency. The DSC benchmark's finding that category-driven routers can direct jailbreaking attempts to weaker, less-safe models while reserving the strongest model for already-simple queries (cf.~Section~\ref{subsec:benchmarks}) is a concrete instance of this asymmetry. Most methods surveyed here collapse this distinction into a single blended score, typically an error rate combined linearly with a cost penalty, as in UniRoute's cost-adjusted scoring or IRT-Router's performance-cost routing score, which cannot separate a costly-but-safe misrouting from a cheap-but-wrong one. A smaller number of methods address this asymmetry more directly, namely FrugalGPT's cost-aware escalation threshold \citep{Chen2024-FrugalGPT} and BEST-Route's user-set thresholds \citep{Ding2025-BEST-Route}, which let operators tune how conservatively the system escalates, and Firewall Routing \citep{Peng2025-FirewallRoutingLLMs}, which specifically targets queries that no model in the pool can answer correctly, where escalation is pure waste rather than a genuine quality-cost trade-off. CRE-Router \citep{Moslem2026-CRE-Router} shows that these two approaches are not mutually exclusive. Its Stage 1 cluster-to-model assignment blends cost and correctness into a single score like UniRoute and IRT-Router, while its Stage 2 quality-estimation classifier adds a direct per-query accept-or-escalate decision like FrugalGPT and BEST-Route, catching individual misroutes that cluster-level assignment alone cannot correct. Such per-query escalation matters most in higher-stakes settings, where an incorrect response carries a real cost beyond the latency or price of escalating, while still keeping the system within its overall cost or latency budget.

\paragraph{Routing collapse.} \citet{Lai2026-RoutingCollapse} document a failure mode specific to routers that predict a scalar performance score for each candidate model. As the cost budget increases, these routers systematically over-select the most expensive model even when cheaper ones would suffice, a pattern they call routing collapse and confirm across RouterBench, MMR-Bench, and several already-cited routers including EmbedLLM, GraphRouter, and IRT-Router. They trace the cause to an objective-decision mismatch, since the router is trained to predict scalar scores but the routing decision is a discrete comparison between near-tied candidates, so small prediction errors flip the outcome and increasingly favour the strongest model as more models become budget-feasible.

%% file: tables/evaluation.tex
\newcolumntype{L}[1]{>{\raggedright\arraybackslash}p{#1}}

\begin{table}[htbp]
\centering
\footnotesize
\begin{tabular}{
>{\raggedright\arraybackslash}p{2.3cm}
>{\raggedright\arraybackslash}p{3.0cm}
>{\raggedright\arraybackslash}p{5.9cm}
>{\raggedright\arraybackslash}p{3.6cm}}
\toprule
\textbf{Method} & \textbf{Baseline(s)} & \textbf{Performance / Cost} & \textbf{Setup} \\
\midrule
Cascade Routing &
Linear interpolation, Routing, Cascade (baseline) &
87.24\% AUC vs.\ 84.48\% for the strongest baseline (11-model, low-noise) &
3/5/11-model subsets tested; monetary cost from RouterBench \\
\addlinespace
GreenServ &
Contextual $\epsilon$-greedy, Thompson sampling &
Best accuracy (71.7\% avg.) but not best AIQ (0.607 vs.\ 0.637 for $\epsilon$-greedy) &
$\sim$36k queries, 9 tasks; 16-LLM pool \\
\addlinespace
IRT-Router &
Small/Large LLM only, HybridLLM, RouteLLM, RouterBench's Predictive Router &
80.69\% accuracy at \$0.55/query vs.\ 80.01\% at \$1.15 for RouterBench's Predictive Router &
Small=Ministral-8B, Large=GPT-4o; cost=\$0.2/M output tokens \\
\addlinespace
MixLLM &
AutoMix, RouteLLM, Zooter, FORC, OptLLM, MetaLLM, Oracle &
97.25\% of GPT-4 quality at 24.18\% cost vs.\ 96.39\% at 32.94\% for best baseline (OptLLM) &
RouterBench + Llama 3.1 8B/70B; 80/20 split \\
\addlinespace
PILOT &
All-to-one, LinUCB, Epoch-Greedy, HybridLLM &
93\% of GPT-4 performance at 25\% of its cost, multi-task RouterBench setting (a distinct single-task MMLU-only result reaches 86\% at 27\% of cost) &
11 LLMs (6 open-source, 5 proprietary) \\
\addlinespace
UniRoute &
ZeroRouter, K-NN, MLP/Matrix Factorization (oracle) &
Two UniRoute variants reported: LearnedMap (0.711, the paper's primary method) and K-Means (0.712, a simpler variant), both vs.\ 0.707 (K-NN), 0.689 (ZeroRouter); oracle baselines higher (0.720--0.723) &
Random 50/50 train/test split of 11 LLMs \\
\bottomrule
\end{tabular}
\caption{Representative methods evaluated on \textbf{RouterBench} (Hu et al., 2024), with baselines, results, and experimental setup as reported by each work.}
\label{tab:routerbench}
\end{table}

\begin{table}[htbp]
\centering
\small
\begin{tabular}{
>{\raggedright\arraybackslash}p{2.3cm}
>{\raggedright\arraybackslash}p{3.0cm}
>{\raggedright\arraybackslash}p{5.9cm}
>{\raggedright\arraybackslash}p{3.6cm}}
\toprule
\textbf{Method} & \textbf{Baseline(s)} & \textbf{Performance / Cost} & \textbf{Setup} \\
\midrule
BEST-Route &
N-label, N-class, clustering-based routing (all from Srivatsa et al., 2024) &
1.59\% quality drop at 60\% cost reduction vs.\ 5.89\% for N-label; N-class/clustering never reduce cost &
8-LLM pool; GPT-4o reference; out-of-distribution evaluation \\
\addlinespace
Confidence-Driven Router &
Random router, TO-Router, RouteLLM &
CPT(80\%)=55.61 vs.\ 78.55 for random; \$3.74 vs.\ \$4.06 at that operating point &
Strong=GPT-4, Weak=Mixtral-8x7B; cost=OpenAI API USD \\
\addlinespace
RouteLLM &
Random router, BERT, Causal LLM, SW Ranking, Matrix Factorization &
Best variant (Matrix Factorization, Arena+Judge-augmented training) needs only 13.40\% of queries routed to GPT-4 to recover 50\% of the GPT-4-vs.-Mixtral quality gap (CPT(50\%)=13.40\%) vs.\ 49.03\% for random routing, a 60.4\% APGR improvement over random &
Strong=gpt-4-1106-preview, Weak=Mixtral 8x7B; cost=\% of GPT-4 calls \\
\addlinespace
Zooter &
BMA (best single model on average), RMR (reward-model ranking; 6 reward-model variants) &
7.11/10 vs.\ 6.72 for BMA; infers 1 of 6 candidates vs.\ RMR's 6 &
6x13B LLAMA-based pool; GPT-4 judge \\
\bottomrule
\end{tabular}
\caption{Representative methods evaluated on \textbf{MT-Bench} (Zheng et al., 2023), with baselines, results, and experimental setup as reported by each work.}
\label{tab:mtbench}
\end{table}

\begin{table}[ht]
\centering
\small
\begin{tabular}{
>{\raggedright\arraybackslash}p{2.3cm}
>{\raggedright\arraybackslash}p{3.0cm}
>{\raggedright\arraybackslash}p{5.9cm}
>{\raggedright\arraybackslash}p{3.6cm}}
\toprule
\textbf{Method} & \textbf{Baseline(s)} & \textbf{Performance / Cost} & \textbf{Setup} \\
\midrule
HybridLLM &
All-at-large, all-at-small, random &
At 20\% cost advantage (queries routed to the small model), $\leq$1\% BART-score quality drop for the medium-gap pair (Llama-2-13B$\rightarrow$GPT-3.5-turbo); at 40\% cost advantage, drop rises to \(\sim\)4\% for the same pair (up to 10.3\% for the large-gap pair, FLAN-T5$\rightarrow$Llama-2-13B) --- quality drop grows faster than cost savings at higher advantage levels &
3 model pairs tested (small/medium/large performance gap); cost=\% of queries routed to the small model (latency/FLOPs/energy explicitly excluded as confounded); GPT-3.5-turbo via OpenAI API \\
\addlinespace
LLM-Blender &
11 individual LLMs (best: Open Assistant); 4 metric-oracle upper bounds (BERTScore\slash BLEURT\slash BARTScore\slash GPT-Rank); 5 automatic rerankers (Random\slash MLM-Scoring\slash SimCLS\slash SummaReranker\slash PairRanker-alone) &
GPT-Rank 3.01 (lower is better) vs.\ 3.90 for the best single LLM (Open Assistant) &
11-LLM pool (Open Assistant, Vicuna, Alpaca, Baize, MOSS, ChatGLM, Koala, Dolly V2, MPT, StableLM, Flan-T5) \\
\bottomrule
\end{tabular}
\caption{Representative methods evaluated on \textbf{MixInstruct} (Jiang et al., 2023), with baselines, results, and experimental setup as reported by each work.}
\label{tab:mixinstruct}
\end{table}

%% file: sections/framework.tex
\section{Towards Multidimensional Routing Systems}
\label{sec:framework}

As we outlined in Section \ref{sec:intro-framework}, the methods reviewed in this survey indicate that practical routing systems are increasingly multidimensional. Rather than selecting a single paradigm, production deployments tend to compose mechanisms across paradigms to satisfy heterogeneous constraints on quality, latency, cost, and safety. They combine pre-generation and post-generation decisions, integrate query-level and response-level signals, and couple offline learning with online adaptation.

One way to design such systems in production is to treat routing as a control pipeline with three coordinated stages: (i) an initial low-cost pre-router based on query and model metadata subject to cost constraints, (ii) a post-generation verifier that estimates response quality or uncertainty of efficient/weaker models, and (iii) an escalation policy that decides whether to accept, refine, reject, or defer to stronger models. 
In this sense, cascades are a natural extension in routing systems.

To demonstrate how this compositional logic plays out in practice, we can consider representative approaches we have discussed in this survey. For example, both FrugalGPT and Cascade Routing instantiate all three pipeline stages; an initial model is selected by a pre-router (a fixed, offline-learned list for FrugalGPT, a per-query adaptive choice at every step for Cascade Routing), a quality estimator evaluates the response, and an escalation policy decides whether to accept or defer to a stronger model. Different compositional patterns emerge in bandit-based systems such as MixLLM and PILOT, which couple a pre-generation routing decision with online feedback to continuously refine the policy, effectively collapsing the verifier and escalation stages into an implicit reward signal. Uncertainty-based methods such as \mbox{CP-Router} and Self-REF occupy a middle ground, using post-generation confidence estimates as the escalation trigger without requiring explicit quality estimation models.
CRE-Router (cf.~Section~\ref{sec:cascades}) composes clustering-based routing with a quality-estimation cascade, and is notable for driving both stages from task-correctness labels alone, jointly optimising model selection under an explicit latency budget while recovering accuracy on hard queries without any additional annotation.
As illustrated by our design-space matrix (cf.~Table~\ref{tab:framework}), these examples suggest that the three-stage pipeline is not a rigid architecture but a flexible template whose stages can be combined, collapsed, or reordered depending on deployment constraints.

This multidimensional perspective also motivates clearer method characterization and comparison. In particular, evaluating routing methods should explicitly report: when decisions are made (pre-generation, post-generation, or multi-stage routing), what signals are used (query, model metadata, response, feedback), and how decisions are computed (heuristic, supervised, bandit, RL policy, or combinations thereof). Methods may activate multiple categories simultaneously, and this compositionality should be reflected in both taxonomy and empirical reporting.

In addition to architectural mapping, our design-space matrix (cf.~Table~\ref{tab:framework}) also reveals structural gaps in the current literature that point to concrete directions for future work. First, no current method simultaneously pairs response-level signals with online adaptation, as uncertainty-based approaches and cascades exploit response signals but remain static once deployed, while bandit-based methods adapt online but operate on query-level signals alone. This suggests an opportunity for bandit-style escalation policies that refine thresholds from deployment feedback. Second, reinforcement learning is limited in the cascading paradigm, and combining learned escalation policies with multi-stage architectures remains largely unexplored. Third, while some systems balance quality, cost, and latency via fixed tradeoff parameters or constraints, few formulate and solve a unified multi-objective optimization problem that treats all metrics as first-class, tunable objectives. Addressing these gaps will be critical for building routing systems that fully exploit the rich design space revealed by the multidimensional framework.

%% file: sections/future.tex
\section{Conclusion and Future Directions}
\label{sec:future}

This survey has covered state-of-the-art approaches to dynamic routing methods for multi-LLM deployment at inference time. We organized approaches into six main paradigms: difficulty-aware routing, human preference alignment, clustering-based methods, reinforcement learning approaches, uncertainty quantification, and cascading systems.
Additionally, we examined emerging work on multimodal routing and discussed evaluation frameworks, including benchmarks and metrics. The key insight is that effective routing systems can outperform even the strongest individual models by strategically leveraging model complementarity and specialization while achieving considerable efficiency gains.

Despite rapid progress, various open challenges remain, with the following potential research directions:
\begin{itemize}[itemsep=2pt, topsep=1.5pt, parsep=1.8pt]
    \item \textbf{Standardized evaluation:} As Section~\ref{sec:evaluation} shows, methods sharing the same benchmark rarely share a metric, model pool, or cost basis, which limits direct cross-examination across approaches. We hope this survey encourages the community toward shared evaluation protocols that make routing methods more directly comparable and interpretable.
    \item \textbf{Generalization:} Several routing methods are evaluated on a fixed set of LLMs and struggle to generalize to new models, domains, or data distributions. Developing retraining-free approaches that transfer across diverse architectures, tasks, and deployment scenarios remains an open challenge.
    \item \textbf{Multi-stage cascades:} Most work explores one-stage routing rather than cascading systems (cf.~Section~\ref{sec:cascades}). As we outlined in Section~\ref{sec:intro-framework} and Section~\ref{sec:framework}, real-world deployments rarely conform to a single paradigm.  Future research should reflect real-world applications where queries and outputs are processed at multiple levels, aiming to strike a balance between quality and efficiency while also ensuring safety.
    \item \textbf{Multimodality:} As models increasingly span vision, audio, and language, routing research should deeply address modality-specific challenges, including fusing multimodal inputs for routing decisions, handling queries requiring multiple modalities, and accounting for varying computational costs of cross-modal architectures.
\end{itemize}

As the landscape of large foundation models continues to diversify, intelligent routing is becoming increasingly critical for efficient deployment. This survey aims at providing researchers and practitioners with a foundation for understanding current approaches and highlighting opportunities for advancing the state of the art in dynamic multi-LLM systems.

%% file: sections/ethics.tex
\section{Ethical Statement}
\label{sec:ethics}

Relying on a single general-purpose LLM structurally underrepresents diverse data, skills, and populations. \citet{Feng2025-MultiLLMCollaboration} identify routing and cascading, alongside other collaboration mechanisms, as ways multiple independently developed models can be composed to address this, since communities that cannot train a frontier model can still contribute a specialized one to the pool.

Nevertheless, routing systems that minimize cost by preferring smaller models will, by design, deliver lower-quality outputs for the queries those models handle least well. If the distribution of difficult queries correlates with underrepresented languages, specialized domains, or non-standard writing styles, then cost-optimized routing could quietly introduce quality disparities across user populations without any individual decision reflecting that intent, and speakers of those languages would also bear a higher routing cost, a disparity already observed in other technical contexts such as tokenization efficiency across languages. The feedback-adaptive methods (bandit-based approaches in particular) add a second concern; routing policies that update from deployment feedback can entrench early distributional biases over time. Acknowledging these disparities is a necessary first step toward addressing them.

Beyond disparities across user populations, routing and cascading can also work against the interests of users generally, even without disparate impact. \citet{Mahmood2025-RoutingCascadesUserChoice} formalize routing and cascading as a Stackelberg game between a provider and a user who can re-prompt or abandon a task. They show that when the provider's cost ranking of models diverges from the user's utility ranking, the provider-optimal policy is not the user-optimal one. In the most severe case, the provider can be incentivized to inflate model latency itself, since discouraging repeated use lowers its own service costs at the expense of user welfare. This is a further mechanism by which a cost-optimized routing system can act against user interests without any single decision reflecting that intent.

Finally, training a routing system can itself be resource-intensive. Reinforcement-learning-based routers, for instance, are more computationally expensive to train than approaches that rely on a small classifier or clusterer. Serving a routing system also requires keeping multiple large models loaded in memory, awaiting possible selection. Incorporating complementary efficiency techniques for managing the underlying model pool is therefore important for routing to deliver its intended efficiency gains.

%% file: sections/acks.tex
\subsubsection*{Acknowledgements}

We sincerely thank Research Ireland's ADAPT Centre and Huawei Research for supporting this work. Special thanks to David Lynch, Magdalena Kacmajor, Vasudevan Nedumpozhimana, and Ammar Abbas for their feedback during the project.

%% file: tables/comparison.tex
\begin{table*}[p]
\tiny
\centering
\resizebox{\textwidth}{!}{%
\begin{tabular}{llll}

\toprule
\textbf{Method} & \textbf{Paradigm} & \textbf{Routing Model / Mechanism} & \textbf{Training Data} \\
\midrule

\textbf{BEST-Route}~{\tiny[§\ref{sec:difficulty}]} \\ {\tiny\citep{Ding2025-BEST-Route}} & Difficulty & DeBERTa-v3-small (multi-head) & Question--score pairs \\
\addlinespace

\textbf{Semantic Router}~{\tiny[§\ref{sec:difficulty}]} \\ {\tiny\citep{Wang2025-vLLM-SemanticRouter}} & Difficulty & ModernBERT classifier & Query intent labels \\
\addlinespace

\textbf{RouteLMT}~{\tiny[§\ref{sec:difficulty}]} \\ {\tiny\citep{Luo2026-RouteLMT}} & Difficulty & LoRA probe on small translator representations & Parallel translation pairs (ComMT) \\
\addlinespace

\textbf{EmbedLLM}~{\tiny[§\ref{sec:difficulty}]} \\ {\tiny\citep{Zhuang2024-EmbedLLM}} & Difficulty & Matrix factorization (model embeddings) & Question-answer correctness data \\
\addlinespace

\textbf{ICL-Router}~{\tiny[§\ref{sec:difficulty}]} \\ {\tiny\citep{Wang2026-ICL-Router}} & Difficulty & LLM router + projector + in-context capability vectors & Query-LLM performance pairs \\
\addlinespace

\textbf{GraphRouter}~{\tiny[§\ref{sec:difficulty}]} \\ {\tiny\citep{Feng2024-GraphRouter}} & Difficulty & Graph Neural Network (GNN) & Historical performance + cost data \\
\addlinespace

\textbf{IRT-Router}~{\tiny[§\ref{sec:difficulty}]} \\ {\tiny\citep{Song2025-IRT-Router}} & Difficulty & IRT-based performance predictor (MIRT/NIRT) & Query-LLM performance pairs \\
\addlinespace

\textbf{RouteLLM}~{\tiny[§\ref{sec:preference}]} \\ {\tiny\citep{Ong2025-RouteLLM}} & Preference & Matrix factorization / BERT / Causal LLM & Chatbot Arena + LLM judge \\
\addlinespace

\textbf{Arch-Router}~{\tiny[§\ref{sec:preference}]} \\ {\tiny\citep{Tran2025-Arch-Router}} & Preference & 1.5B parameter LLM & Synthetic preference data \\
\addlinespace

\textbf{Hybrid LLM}~{\tiny[§\ref{sec:preference}]} \\ {\tiny\citep{Ding2024-HybridLLM}} & Preference & LLM-based classifier & Synthetic preference data \\
\addlinespace

\textbf{P2L}~{\tiny[§\ref{sec:preference}]} \\ {\tiny\citep{Frick2025-Prompt-to-Leaderboard}} & Preference & LLM (Bradley-Terry coefficients) & Chatbot Arena preferences \\
\addlinespace

\textbf{Eagle}~{\tiny[§\ref{sec:preference}]} \\ {\tiny\citep{Zhao2024-Eagle}} & Preference & ELO ranking (training-free) & Historical pairwise comparisons \\
\addlinespace

\textbf{Zooter}~{\tiny[§\ref{sec:preference}]} \\ {\tiny\citep{Lu2024-Zooter}} & Preference & mDeBERTa-v3-base classifier & Reward model preference labels \\
\addlinespace

\textbf{UniRoute}~{\tiny[§\ref{sec:clustering}]} \\ {\tiny\citep{Jitkrittum2026-UniRoute}} & Clustering & K-means + unsupervised text embedder & Unlabelled queries + validation set \\
\addlinespace

\textbf{Avengers-Pro}~{\tiny[§\ref{sec:clustering}]} \\ {\tiny\citep{Zhang2025-Avengers-Pro}} & Clustering & K-means + text embedder + scoring & Labelled query-answer pairs \\
\addlinespace

\textbf{Router-R1}~{\tiny[§\ref{sec:rl}]} \\ {\tiny\citep{Zhang2025-Router-R1}} & RL (Policy Opt.) & Qwen2.5-3B / LLaMA-3.2-3B (PPO) & Question--score pairs \\
\addlinespace

\textbf{R2-Reasoner}~{\tiny[§\ref{sec:rl}]} \\ {\tiny\citep{Shao2025-R2-Reasoner}} & RL (Policy Opt.) & Task Decomposer + Subtask Allocator (GRPO) & SFT + RL training data \\
\addlinespace

\textbf{MetaLLM}~{\tiny[§\ref{sec:rl}]} \\ {\tiny\citep{Nguyen2025-MetaLLM}} & RL (Bandit) & Multi-armed bandit & Online accuracy-cost feedback \\
\addlinespace

\textbf{MixLLM}~{\tiny[§\ref{sec:rl}]} \\ {\tiny\citep{Wang2025-MixLLM}} & RL (Bandit) & Contextual bandit + policy gradient & Binary user feedback (online) \\
\addlinespace

\textbf{PILOT}~{\tiny[§\ref{sec:rl}]} \\ {\tiny\citep{Panda2025-PILOT}} & RL (Bandit) & LinUCB contextual bandit & Chatbot Arena priors + online feedback \\
\addlinespace

\textbf{GreenServ}~{\tiny[§\ref{sec:rl}]} \\ {\tiny\citep{Ziller2026-GreenServ}} & RL (Bandit) & LinUCB contextual bandit & Online energy consumption + accuracy feedback \\
\addlinespace

\textbf{Dueling Feedback}~{\tiny[§\ref{sec:rl}]} \\ {\tiny\citep{Chiang2025-Dueling-Feedback}} & RL (Bandit) & CCFT text encoder + FGTS.CDB & Pairwise preference comparisons \\
\addlinespace

\textbf{TI-UCB}~{\tiny[§\ref{sec:rl}]} \\ {\tiny\citep{Xia2024-TI-UCB}} & RL (Bandit) & Time-increasing UCB bandit & Online model performance \\
\addlinespace

\textbf{CP-Router}~{\tiny[§\ref{sec:uncertainty}]} \\ {\tiny\citep{Su2025-CP-Router}} & Uncertainty & Conformal prediction on logits & Calibration set \\
\addlinespace

\textbf{Confidence-Driven}~{\tiny[§\ref{sec:uncertainty}]} \\ {\tiny\citep{Zhang2025-LLMJudge-UncertaintyRouting}} & Uncertainty & Semantic entropy (entailment clustering) + classifier & SE-derived preference labels \\
\addlinespace

\textbf{AutoMix}~{\tiny[§\ref{sec:cascades}]} \\ {\tiny\citep{Aggarwal2024-AutoMix}} & Uncertainty (Cascade) & POMDP router + few-shot self-verification & Few-shot examples (no fine-tuning) \\
\addlinespace

\textbf{Self-REF}~{\tiny[§\ref{sec:cascades}]} \\ {\tiny\citep{Chuang2025-Self-REF}} & Uncertainty (Cascade) & Fine-tuned LLM w/ confidence tokens & Correctness-labelled training data \\
\addlinespace

\textbf{FrugalGPT}~{\tiny[§\ref{sec:cascades}]} \\ {\tiny\citep{Chen2024-FrugalGPT}} & QE (Cascade) & LLM router + DistilBERT quality estimator & Query-response pairs \\
\addlinespace

\textbf{Cascade Routing}~{\tiny[§\ref{sec:cascades}]} \\ {\tiny\citep{Dekoninck2025-CascadeRouting}} & QE (Cascade) & Iterative model selection + quality estimator & Query-response pairs (log probabilities) \\
\addlinespace

\textbf{CRE-Router}~{\tiny[§\ref{sec:cascades}]} \\ {\tiny\citep{Moslem2026-CRE-Router}} & Clustering \& QE (Cascade) & K-means clustering + quality-estimation classifier & Task-correctness labels \\
\addlinespace

\textbf{LLM-Blender}~{\tiny[§\ref{sec:cascades}]} \\ {\tiny\citep{Jiang2023-LLM-Blender}} & Ensemble (Cascade) & Cross-attention Pair Ranker + Gen Fuser & Pairwise preference labels \\
\addlinespace

\bottomrule
\end{tabular}
}
\caption{Summary of LLM routing and cascading methods, reflecting the main paradigms of this survey.}
\label{tab:methods}
\end{table*}